\documentclass[twocolumn]{aastex62}

\graphicspath{{./}{figures/}}

\received{January 1, 2018}
\revised{January 7, 2018}
\accepted{\today}

\submitjournal{ApJ}

\usepackage{multirow,hhline,graphicx,array}
\newcolumntype{M}[1]{>{\centering\arraybackslash}m{#1}}
\usepackage{rotating} 

\shorttitle{Multiwavelength Absolute Magnitudes of RC}
\shortauthors{Plevne et al. 2020}

\begin{document}

\title{Multiwavelength Absolute Magnitudes and Colours of Red Clump Stars in {\it Gaia} Era  \footnote{Released on January, 8th, 2018}}

\correspondingauthor{Olcay Plevne}
\email{olcayplevne@istanbul.edu.tr}

\author[0000-0002-0786-7307]{Olcay Plevne}
\affil{Istanbul University, Graduate School of Science, Department of Astronomy and Space Sciences, \\
34119, Beyaz{\i}t, Istanbul, TURKEY \\}
\affil{Istanbul University, Faculty of Science, Department of Astronomy  and Space Sciences \\
34119, Beyaz{\i}t, Istanbul, TURKEY }

\author{\"Ozgecan \"Onal Ta\c s}
\author{Sel\c cuk Bilir}
\affil{Istanbul University, Faculty of Science, Department of Astronomy  and Space Sciences \\
34119, Beyaz{\i}t, Istanbul, TURKEY }

\author{George M. Seabroke}
\affil{Mullard Space Science Laboratory, University College London, Holmbury St Mary, Dorking, RH5 6NT, UNITED KINGDOM}

\begin{abstract}
This study presents the multi-wavelength investigation of the absolute magnitudes and colours of the red clump (RC) stars selected from APOGEE and GALAH DR2 combined catalogue which is complemented with {\it Gaia} DR2 astrometric data and multi-wavelength photometric data of {\it GALEX} GR6/7, SDSS DR7, {\it Gaia} DR2, 2MASS and {\it WISE} sky surveys. The analyses are centred on the different distance estimation methods using {\it Gaia} trigonometric parallaxes, (1/$\varpi$) and Bayes statistics, and chemically defined Galactic disc populations on [$\alpha$/Fe]$\times$[Fe/H] plane. Such investigation questions the long studied problem of the population effects on RC luminosity. Using two different distance estimation approach, (i) chemical thin and chemical thick disc RC stars are shown to have different absolute magnitudes, while colours remain the same in all photometric bands. Absolute magnitudes vary between -0.12 and +0.13 mag for the 1/$\varpi$ with the change of the Galactic population. This variation in absolute magnitudes is found to be larger for the other method. (ii) The Besan\c con population synthesis model of Galaxy for 2MASS photometry, in which the absolute magnitude difference between chemical populations were found between -0.35 and -0.40 mag from thin disc to thick disc. When results compared with each other, differences of absolute magnitudes are about three times larger in the model than observations. We confirm that the RC absolute magnitudes depend on $\alpha$-element abundances of Galactic populations.
\end{abstract}



\section{Introduction}
Galactic archaeology studies require objects that are observable in a vast space volume, like the Gaiasphere, in order to construct formation and evolution scenarios that lead to the present day picture from the past events of the Milky Way galaxy. The critical parameter for these studies is reliably estimated distance, which is the current hot topic. As of the {\it Gaia} \citep{GC16} era started, our blurred vision of Galactic structure is become more clear starting from the Solar neighbourhood, with the support of the advanced computer programming abilities that grow within the scientific community. The second data release of {\it Gaia} \citep{GC18} contains the largest data set with the most accurate 5D astrometric data, superseding its predecessor, the {\it Hipparcos} astrometry satellite \citep{Perryman97}, by combining spectroscopic and astrometric data allows to obtain the most accurate 12-D parameter space for Galactic evolution studies. The 25 year gap between {\it Hipparcos} and {\it Gaia} was filled with developing and/or improving alternative distance estimation methods via photometric and/or spectroscopic properties of various celestial objects for different distance ranges in the Universe, so called the distance ladder.  Cepheid variables and Type Ia supernovae, which are considered as ``standard candles'' due to their absolute magnitudes at a certain evolutionary stage, which they radiate precise energy output. Amongst these, one other object is alternatively considered to be a standard candle in the last forty years, are {\it red clump stars}. 

Red clump (RC) stars, which are visible throughout great distances such as neighbouring galaxies and their almost constant brightness give them a potential standard candle status. All the data regarding their standard candle status were obtained from spectroscopic survey data with spectro-photometric distances before the {\it Gaia} era.

\citet{CL69} showed that it is possible to determine clumps of red giant stars from their colour and surface gravity properties and also their distances can be calculated using photometric data. Investigating the red giant stars in Hertzsprung-Russell (HR) diagrams of rich and old open clusters, \citet{Cannon70} pointed out that these red giant stars are clumped around $(M_V,B-V)$=(1,1) mag on the colour-magnitude diagram. Then, based on the stellar interior models of the time,  \citet{Faulkner66} and \citet{Iben67b}, the possible cause of this clumping is examined and arrive the conclusion that these are the post-helium flash stars with stable helium burning inside their cores and their masses are less than 2.25$M_{\odot}$. According to theory, RCs are metal rich, low-mass stars ($M<2.25M_{\odot}$) with stable helium burning core, which cause a narrow range of luminosity. Since their first recognition by \citet{Cannon70}, there have been numerous debates on intrinsic properties, populations and absolute magnitudes of RC stars in different parts of the electromagnetic spectrum, i.e. ultra-violet (UV), optical, near-infrared (NIR) and mid-infrared (MIR).

A significant contribution to RC studies started with the more precise parallax measurements from {\it Hipparcos} mission \citep{Perryman98}. RC region was a very prominent feature on the colour-magnitude diagram of the Solar neighbourhood stars observed by {\it Hipparcos}. The {\it Hipparcos} catalogue \citep{ESA97} contained approximately 600 RC stars with relative parallax errors less than 0.1 \citep{Girardi98}. These precise parallax measurements triggered extensive studies on the dependencies of the RC absolute magnitude to age and chemistry. By doing so, the absolute magnitudes of RC stars selected from different regions of the Milky Way and neighbouring galaxies were calibrated many times. First studies used optical photometric bands $V$ and $I$. Metallicity dependence of the $I$-band magnitudes were investigated  by several authors in the literature \citep{PS98, Perryman98, Udalski00}, and it is found weak for this region. Moreover, \citet{Sarajedini99} and \cite{Twarog99} claimed the RC magnitudes depend on the metallicity and age by using RCs in open clusters, which is supported by the models later on \citep{GS01, SG02}. Using {\it Hipparcos} stars \citet{Alves00} made a calibration for 2MASS $K_s$-band \citep{Cutri03}, which is less affected by reddening and mild systematic dependence on metallicity. Under the assumption of no reddening \citet{Alves00} found $M_{K_s} =-1.61\pm0.03$ mag, with a linear relation with metallicity, i.e. $M_{K_s}= 0.57(0.36)\times{\rm [Fe/H]}-1.64(0.07)$. 

The real leap on RC studies occurred when the re-reduced {\it Hipparcos} data \citep{vanLeeuwen07}, in which the parallax measurements and their respective errors were updated. \citet{Groenewegen08} calculated the absolute magnitudes for a sample selected from re-reduced {\it Hipparcos} catalogue and found mean absolute magnitudes in optical $I$ and NIR $K_s$ bands as $\langle M_I \rangle=-0.22\pm0.03$ mag and $\langle M_{K_s}\rangle=-1.54\pm0.04$ mag, respectively. Also, \citet{Groenewegen08} modelled a synthetic RC star sample by applying a number of selection criteria on metallicity, age, magnitude range etc. in order to infer the effects of population and absolute magnitude selection. These analyses showed that the $M_I$ and $M_{K_s}$ are weakly related to metallicity and $V-K$ colour, respectively. As the wide area NIR all-sky surveys 2MASS, DENIS became more complete the RC studies to validate the standard candle status are spread out to the NIR photometric bands. According to \citet{Salaris13}, $K_s$-band is the ideal photometric band to determine RC distances, because the star formation history does not change in this band. In \cite{Bilir13a}, metallicity dependence of RC stars are selected from open and globular clusters in optical bands. The first absolute magnitude determination in {\it WISE} photometric bands was performed by \citet{YG13}'s large RC sample, and there have been numerous studies \citep{Chen17, Ruiz18}. In a recent study, \citet{Mohammed19} analysed APOGEE DR14 RC stars in {\it GALEX} $NUV$-band with {\it Gaia} $G$-band, and found a strong dependence of colour on effective temperature and metallicity.

Chemical abundances of stars changes drastically in Galaxy-wide distances so that various metallicity gradients were obtained in various radial and vertical directions from the Galactic centre and the Galactic plane, respectively \citep{Coskunoglu12, Plevne15, Onaltas16, Onaltas18, Tuncel19}. Up until the {\it Gaia} era, our RC knowledge was gathered from the stars within the Solar neighbourhood with different radii ($5< R <10$ kpc and $-2 < |Z|< 2$ kpc). Nowadays, thanks to the precise astrometry of {\it Gaia} DR2 \citep{GC18} and high resolution ($R > 22,000$) and high signal-to-noise ($S/N > 100$) of APOGEE \citep{Majewski17} and GALAH DR2 \citep{Buder18} spectroscopy, RC stars became appealing objects to study.

Many studies in the last 25 years established various median absolute magnitudes in optical, NIR and MIR parts of the electromagnetic spectrum \citep{Girardi16}. However, these studies were biased towards the bright nearby objects within the Solar neighbourhood due to sample selection using relative parallax errors to obtain dependable sub-samples. According to the review study of \citet{Girardi16}, RC stars are more abundant than horizontal branch stars. Thus, these stars are one of the main targets of brightness--limited sky surveys. Based-on their spectral types (G8-K2), they are ideal objects for accurate radial velocity and chemical abundance determination \citep{Saguner11}. 

The era of wide-area imaging surveys such as {\it GALEX}  \citep{Martin05}, SDSS \citep{Abazijan04}, 2MASS \citep{Skrutskie06} and {\it WISE} \citep{Wright10}, allow probing the multi-wavelength properties (from UV to MIR) of any selected stellar population. Especially the combination of these surveys with the precise astrometric properties of {\it Gaia} DR2 \citep{GC18} and high resolution and high $S/N$ spectroscopic survey like APOGEE \citep{Allende08} and GALAH \citep{dS16} allow an in depth study. 

This study is focused on the multi-wavelength absolute magnitudes and colours of RC stars in low- and high-[$\alpha$/Fe] populations in the Galactic disc. The study also deals with different distance estimation methods. The paper is organised as follows. Data selection and RC identification is described in Sect. 2, distance estimation methods, interstellar reddening and chemical separation of RC stars are presented in Sect. 3, and results of the multi-wavelength absolute magnitudes and colours are given in Sect. 4. Results of a mock catalogue are given in Sect. 5, RC contamination and the effects of the absolute magnitudes on estimated distances are given in Sect. 6, and summary and conclusions are given in Sect. 7.

\begin{figure}[!h]
\centering
\includegraphics[width=\columnwidth]{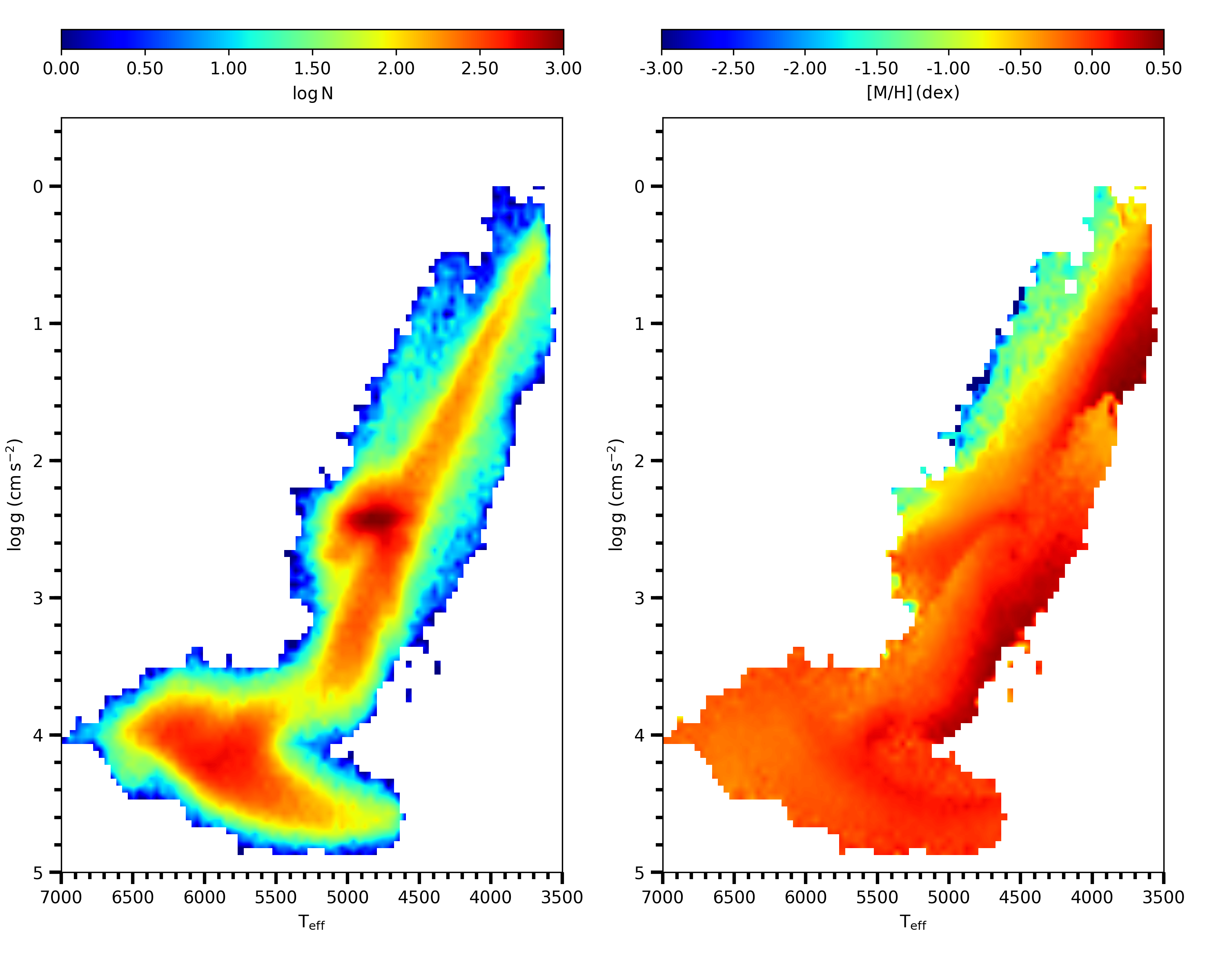}
\caption{HR diagrams of the APOGEE and GALAH DR2 combined catalogue stars, which are colour coded for the stellar number density (left panel) and the metallicity (right panel), respectively. } 
\label{fig:Fig1}
\end {figure} 

\section{Data} \label{sec:data}
\subsection{Red Clump Star Selection}

\begin{figure}[!h]
\centering
\includegraphics[width=\columnwidth]{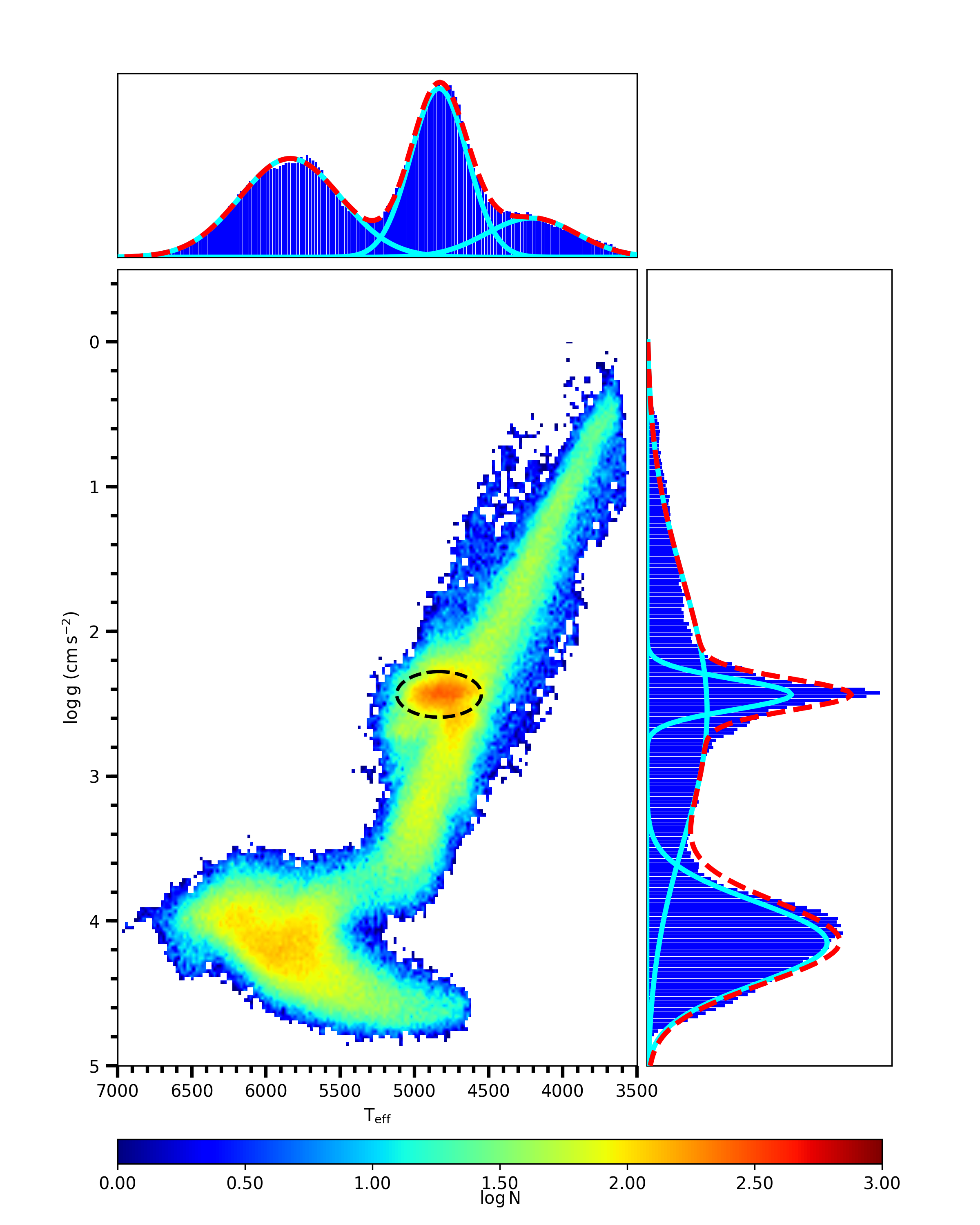}
\caption{Separation of different luminosity classes and selection of the RC stars on HR diagram. Middle panel shows the Kiel diagram of all stars. Upper and right panels give the frequency distributions of $T_{eff}$ and log $g$. Red dashed lines give the overall distribution of stars in each parameter. Turquoise solid lines give the distribution of each luminosity class embedded in the overall distribution.} 
\label{fig:Fig2}
\end {figure} 

In this study, we use spectroscopic data provided by SDSS-IV DR14 APOGEE \citep{Majewski17} and GALAH DR2 \citep{Buder18} surveys. Both surveys have similarities such as high resolution, high $S/N$, spectral analysis pipeline {\it The Cannon} \citep{Ness15}. The only difference between these surveys is their observation grounds, one is made from the Apache Point Observatory in the Northern hemisphere, and the other is made at the Anglo Australian Observatory in the Southern hemisphere. Preliminary data selection is made by eliminating the samples from the objects with low $S/N$ spectra . Then, stars with missing $T_{eff}$, $\log g$, [Fe/H], [$\alpha$/Fe] and radial velocity data are also eliminated. Moreover, selection $flagcannon$=0 and observation with the highest $S/N$ are selected. As a result of these cuts, 154,801 stars in APOGEE and 188,750 stars in GALAH DR2 are found. By combining both catalogues, a master catalogue of 343,551 stars is obtained. This catalogue is named {\it APOGEE-GALAH Red Clump} (AGRC) catalogue.  Spectroscopic HR diagrams of these stars that are plotted by stellar number density ($\log N$) and metallicity ([Fe/H]) are shown in Fig. \ref{fig:Fig1}. There are three more populated regions on the diagram which correspond to main-sequence (MS), red giant branch (RGB) and red clump (RC) stars. Effective  temperature of stars varies between $3500\leq T_{eff}~{\rm (K)}\leq 7000$ and their metallicity varies between $-3\leq {\rm [Fe/H]~(dex)}\leq+0.5$. In order to separate MS, RGB and RC regions, three Gaussian functions are fitted on the stellar number density distribution of each region as shown in Fig. \ref{fig:Fig2}. According to the Gaussian fits MS population covers $4500 < T_{eff}~{\rm (K)} < 7000$ and $3.6< \log g~{\rm (cgs)} < 4.8$, RGB population covers $3500 < T_{eff}~{\rm (K)} < 5300$ and $0 < \log g~{\rm (cgs)} < 3.8$ and RC population covers $4500 < T_{eff}~{\rm (K)} < 5200$ and $2.1 < \log g~{\rm (cgs)} < 2.7$ on HR diagram. Most likely region where RC population resides on the HR diagram has central coordinates ($T_{eff}$, $\log g$) = (4834$\pm$190 K, 2.43$\pm$0.105 cgs), and standard deviations are calculated with the width of half maximum of the Gaussian distributions, i.e. $FWHM=2.35\times\sigma$. RC population is defined by selecting the 2$\sigma$ region around these central coordinates, which is shown as a black short dashed ellipse on the HR diagram. There are 47,537 stars inside this ellipse.

Spectroscopic data are complemented with astrometric data from {\it Gaia} DR2. Quality of trigonometric parallaxes are determined by calculating the relative parallax errors ($\sigma_\varpi/\varpi$) of the sample stars. There exists a systematic scatter in trigonometric parallax measurements, according to \cite{LK73}. Trigonometric parallaxes are affected by bias so that the obtained distances scatter as the observed volume increase. Bias correction function is introduced by \cite{Smith87}'s study, and further studied for {\it Hipparcos} mission by \cite{Oudmaijer02}. Based on the comparative analysis between ground- and space-based parallax measurements with relative parallax errors, it is shown that there is a limiting value of $\sigma_\varpi/\varpi$=0.175. This is defined as the upper limit to correct parallaxes for LK bias. 

\begin{figure}
\centering
\includegraphics[width=\columnwidth]{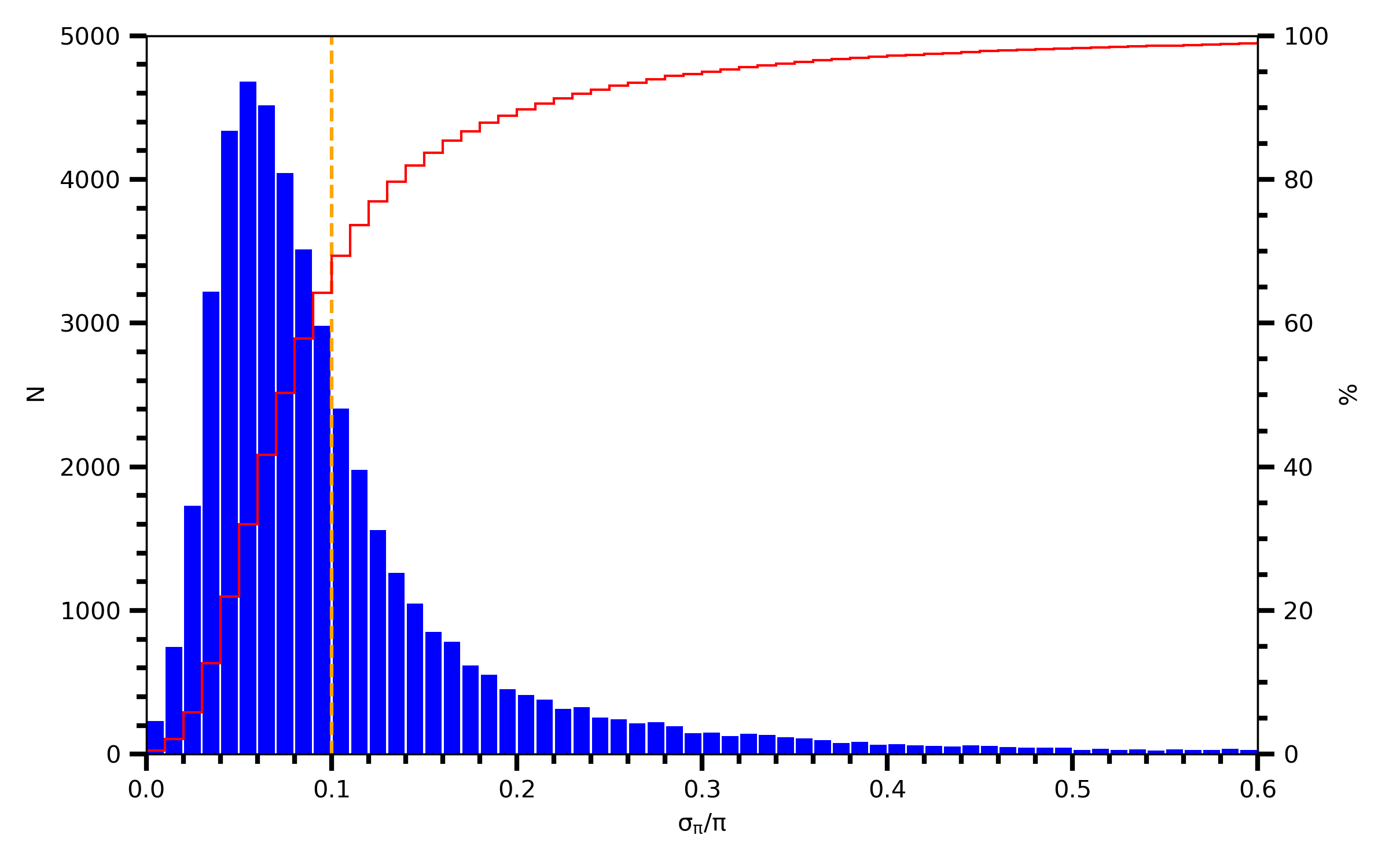}
\caption{Relative parallax error distribution of the AGRC stars. Step function is gives the increase of percentage of the AGRC stars with increasing relative parallax errors. Yellow dashed line shows the limit of $\sigma_{\varpi}/\varpi$=0.1.} 
\label{fig:Fig3}
\end {figure}

Relative parallax distribution of AGRC catalogue is given in Fig. \ref{fig:Fig3}. The relative parallax error distribution range up to 0.6, the median value of AGRC sample is 0.08. In this study, the relative parallax error limit is selected as 0.1 because we wanted to include stars from different Galactic population to the AGRC sample. Then, final sample of AGRC catalogue is 23,880. The AGRC catalogue is separated into relative parallax error sub-sample intervals at 0.05, 0.08, 0.10. These intervals include 35\%, 45\% and 20\% of the AGRC sample. For these sub-samples, The LK corrections to be applied to the {\it Gaia} parallaxes using  Eq. 12 of \citet{Smith87}'s were calculated as $\leq 1\%$, 1\%-2.6\% and 2.6\%-4.2\% respectively. Then, we decided to not to apply LK correction to the AGRC catalogue \citep[see also,][]{Celebi19}.

\begin{figure}[!h]
    \centering
    \includegraphics[width=\columnwidth]{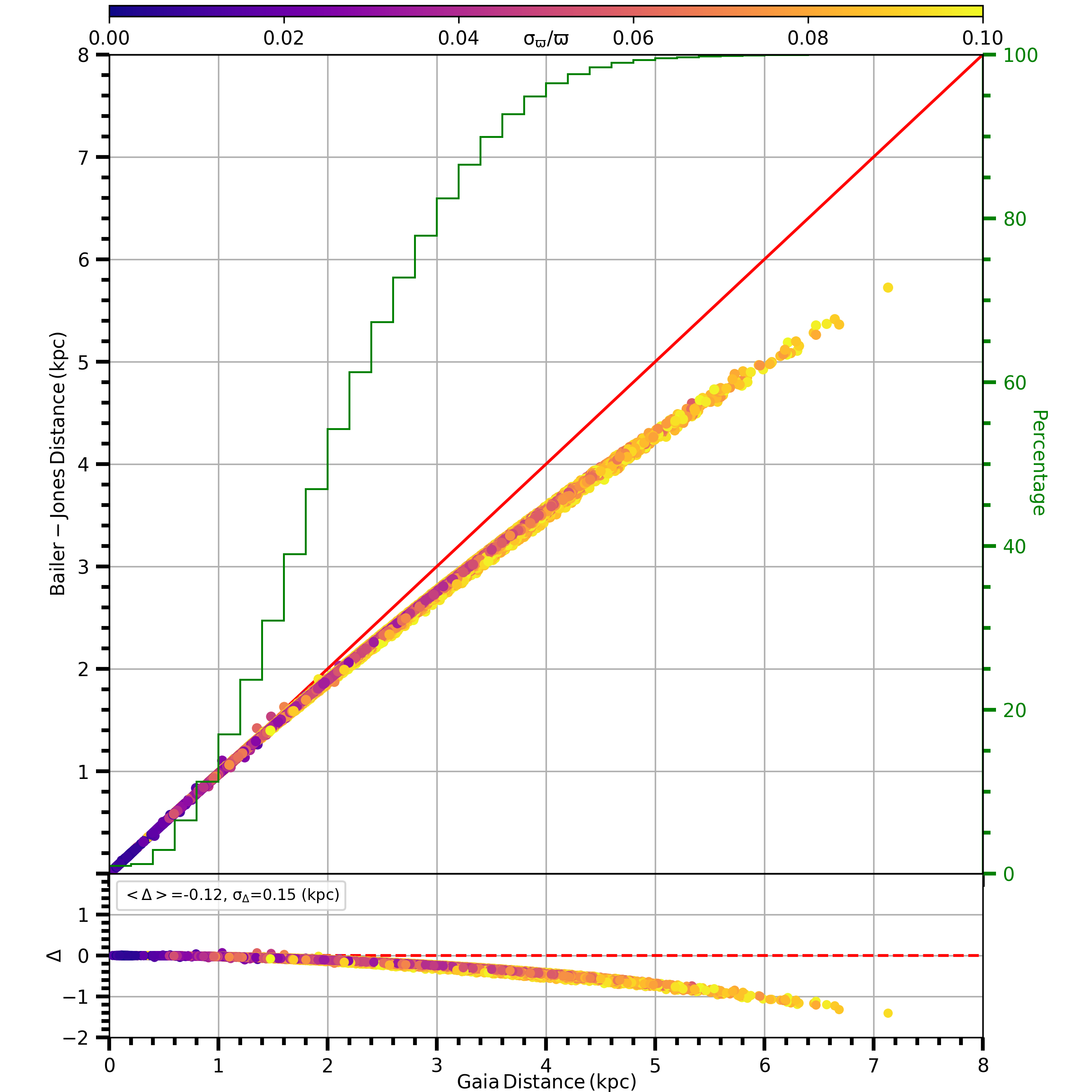}
\caption{Distance comparison between 1/$\varpi$ and BJ18 methods using AGRC catalogue. Stars are represented with colours of individual relative parallax errors. Also, on the right y-axis, and on the background the increment of the sample with increasing distances is shown with green step function. In the lower panel the residual distances are shown with $\langle\Delta\rangle$=0.12 kpc and $\sigma_{\Delta}=0.15$ kpc. Red dashed line represents the zero line.} \label{fig:Fig25}
\end {figure}  

\section{Methods}
\subsection{RC Distances}
RC distances are estimated using two separate methods. One is the conventional inverse-parallax (1/$\varpi$) method, in which the distances are directly calculated from {\it Gaia} DR2 trigonometric parallaxes \citep{GC18}. The other is the probabilistic analysis using priors like trigonometric parallaxes and their uncertainties to estimate the source distances by considering the variations based on the Galactic coordinates and scale height of the disc in a model of the Galaxy \citep{BJ15,BJ18}; we call it the BJ18 method. As it is pointed out by the author, the observational errors in parallaxes can produce bias in samples. This bias is especially effective when the trigonometric parallax values are either negative or their relative errors are larger than 0.2. 

Comparison between stellar distances with $1/\varpi$ and BJ18 methods are shown in Fig. \ref{fig:Fig25}. Data points are coloured based on the relative parallax errors. The increase in percentages of the AGRC stars along with the distance is given in the upper panel, while distance residuals are given in below panel. AGRC catalogue covers the stars within 7 kpc distances and 80\% of the sample lies less than 3 kpc distance, according to Fig. \ref{fig:Fig25}. Distances start to deviate from each other at 1 kpc and this  becomes apparent at 2 kpc. The distance difference becomes larger with increasing distance and reaches 1 kpc at 6 kpc distance in {$1/\varpi$} scale. In this figure, relative parallax errors of RC stars are colour coded, and relative parallax errors increase with the distance. This study implies that the distance estimation method is not a critical ingredient up to 2 kpc.

\begin{figure*}
    \centering
    \includegraphics[scale=0.5]{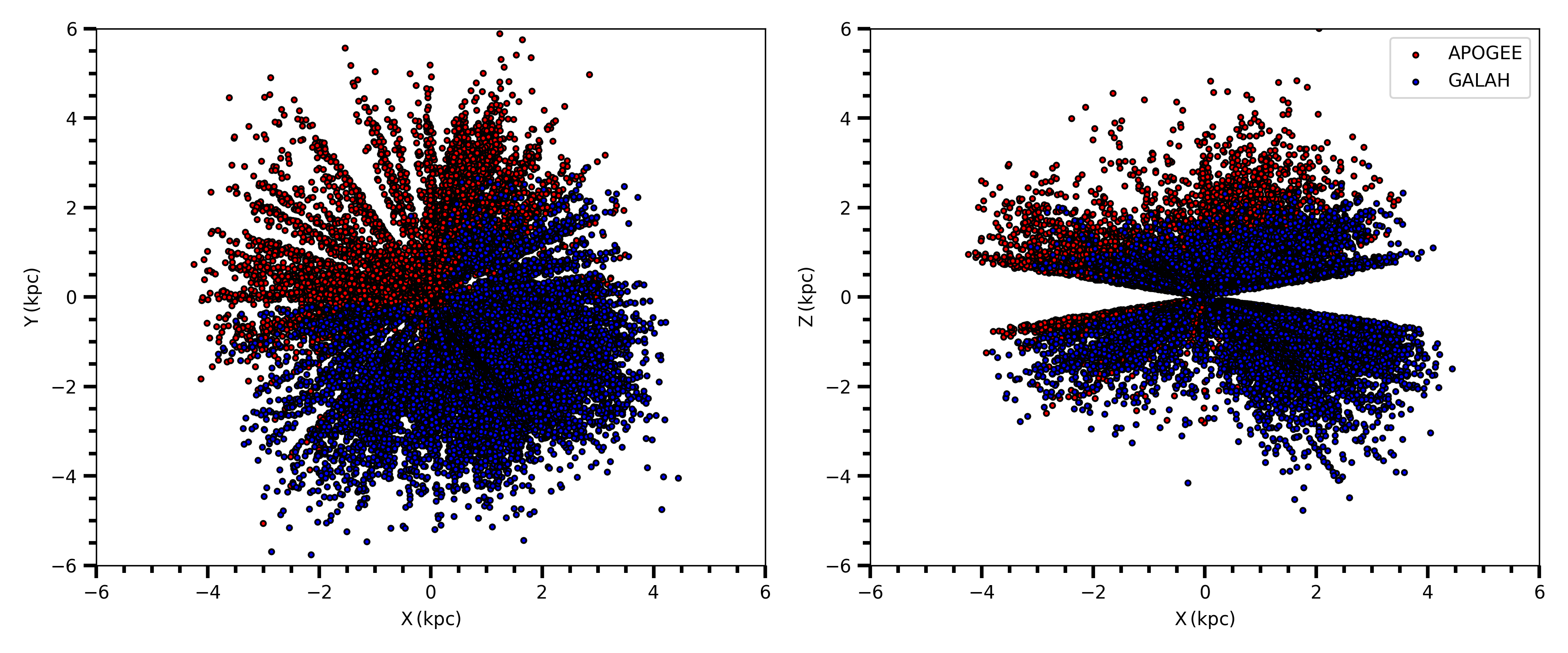}
     \includegraphics[scale=0.6]{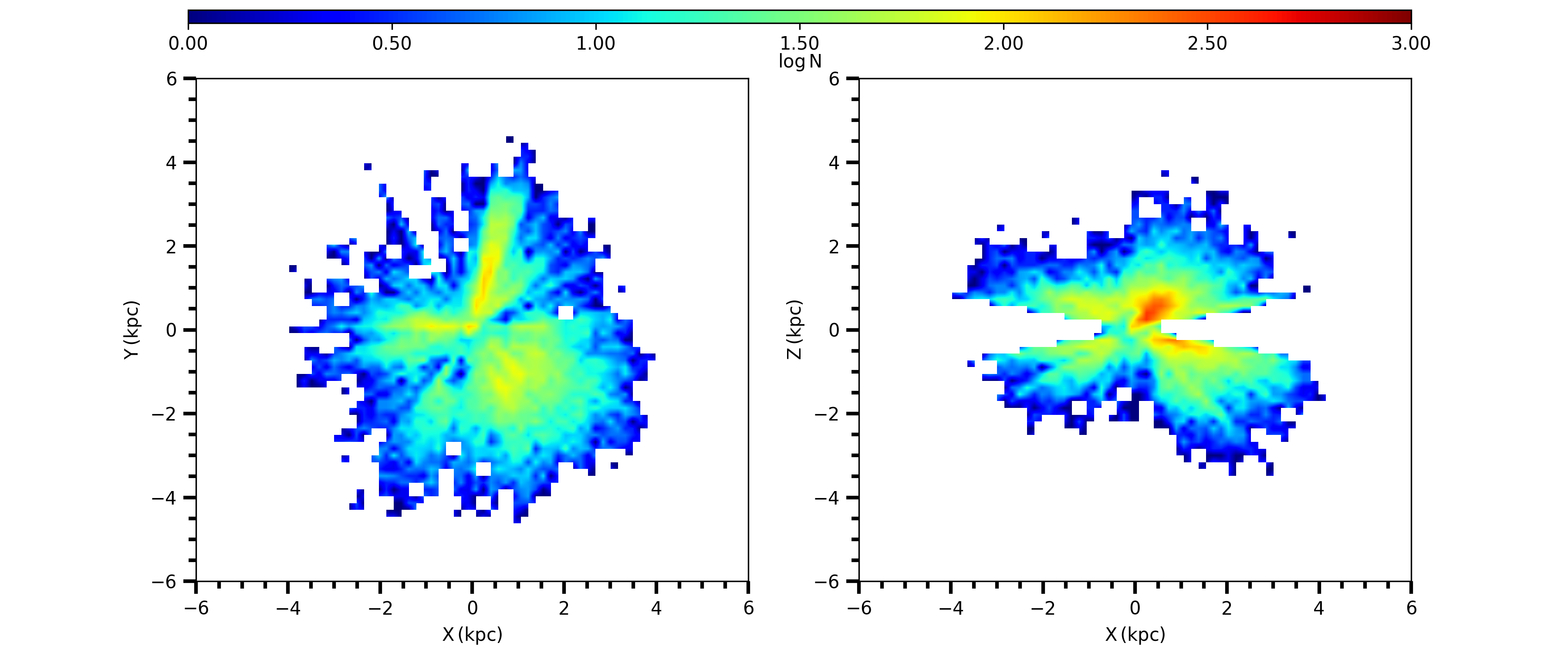}
    \caption{Distribution of RC stars on $X$-$Y$ and $X$-$Z$ planes given by APOGEE (red) and GALAH (blue) surveys (upper panel), and coloured by logarithmic number density (lower panel). The Galactic centre is at approximately +8 kpc on the x-axis \citep{Majewski93}.}
    \label{fig:rc_xyz}
\end{figure*}

Heliocentric coordinates of AGRC stars are calculated using their Galactic coordinates and distances. Heliocentric distance distributions of RC stars on $X$-$Y$ (left panels) and $X$-$Z$ (right panels) planes only for 1/$\varpi$ method is given in Fig. \ref{fig:rc_xyz}. On the upper panels, red and blue circles denote for APOGEE and GALAH DR2 surveys, respectively. In bottom panels, the RC stars are shown in logarithmic number density. Median distances to the Sun and median heliocentric distance components for APOGEE, GALAH DR2 and all sample are listed in Table \ref{table:Table1}.

\begin{table}[h]
\setlength{\tabcolsep}{2pt}
\scriptsize
\centering
\caption{Distance distribution of RC stars for APOGEE, GALAH DR2 and all stars for 1/$\varpi$ (upper panel) and BJ18 (lower panel) methods. Columns give sample, median values of distance, heliocentric $X$, $Y$ and $Z$ distances and number of stars.}
\begin{tabular}{cccccc}
\cline{2-6}
    &\multicolumn{5}{c}{1/$\varpi$ method} \\   
\cline{2-6} 
    & $\tilde{d}$ (kpc) & $\tilde{X}$ (kpc)& $\tilde{Y}$ (kpc) & $\tilde{Z}$ (kpc) & $N$  \\
\cline{1-6} 
APOGEE     &$1.85\pm1.02$  & $0.11\pm1.34$ &  $0.83\pm1.66$  & $0.56\pm1.04$ & 10,798 \\
GALAH      &$2.25\pm0.99$  & $0.96\pm1.24$ & $-1.34\pm1.14$  & $-0.38\pm0.97$ & 13,082 \\
All        &$2.08\pm1.02$  & $0.49\pm1.19$ &  $-0.39\pm1.21$  &  $0.26\pm0.87$ & 23,880 \\
\hline
\hline
    &\multicolumn{5}{c}{BJ18 method} \\ 
\cline{2-6} 
     & $\tilde{d}$ (kpc) & $\tilde{X}$ (kpc) & $\tilde{Y}$ (kpc) & $\tilde{Z}$ (kpc) & $N$\\
\hline
APOGEE     & $1.78\pm0.87$ & $0.12\pm1.10$ & $0.83\pm1.10$  & $0.57\pm0.78$  & 10,798 \\
GALAH      & $2.21\pm0.99$ & $0.97\pm1.32$ & $-1.35\pm1.02$ & $-0.39\pm0.89$ & 13,082 \\
All        & $1.96\pm1.00$ & $0.49\pm1.22$ & $-0.39\pm1.52$ & $0.27\pm0.94$  & 23,880 \\
\hline
\label{table:Table1}
\end{tabular}
\end{table}


\subsection{Chemical Separation of RC Stars with Unsupervised Machine Learning}
Galactic disc, as a dynamically evolving object has regions that are subjected to different internal and external dynamics and chemical processes during its 13 Gyrs evolution. It is well known that kinematic properties of stars can be altered especially in presence of a dominant angular momentum re-distributor (or perturber) such as Galactic bar, spiral arms and giant molecular clouds or even streams of stars. However, chemistry evolves differently in the course of time, depending on the initial mass of stars. Trends in chemistry are suggested by \cite{FBH02} study indicating that stars keep memory of the chemical structure of their birth cluster. Chemical characteristics have been proposed as a way to parameterize the properties of various stellar populations in the Milky Way (scale height, scale length, kinematics, see \cite{Bovy12}). Throughout the Galactic disc, the abundance ratios ([Fe/H], [$\alpha$/Fe] etc.) of stellar samples are known to change towards the radial direction from the Galactic centre or in the vertical direction from the Galactic plane \citep{Bilir06, Bilir08, Bilir12, Cabrera07a}. There is a strong evidence that there are two main structures embedded in the Galactic disc with clearly distinct, but partly overlapping properties, which are known as thin disc and thick disc \citep{GW85, Gratton96, Fuhrmann98, Prochaska00, BFL03, RLAP06, Bensby07, Fuhrmann08, Navarro11, Haywood19, Karaali19}. Thin disc stars are younger, rich in [Fe/H], poor in [$\alpha$/Fe], and have cold kinematic properties, while thick disc stars are significantly older, poor in [Fe/H], rich in [$\alpha$/Fe] and have hot kinematic properties. Distribution of stars on this plane is a reflection of the chemical evolution of the Milky Way. Investigations of the chemical evolution of the Galactic disc have shown that the disc has undergone at least two different periods of formation \citet{CMG97}. The stars that are born during these formation periods can be separated on [$\alpha$/Fe]$\times$[Fe/H] plane. It turns out that the chemical abundances plotted in this plane can be used to disentangle the components of the Galactic disc in age \citep{WG88}. This discrete structure persists even at different radial distances along the Galactic disc \citep{Haywood08}. Although the disc consists of two separate chemical populations, different studies examining this distinction have not been able to establish a specific criterion for separation because populations are intertwined in the chemical plane.

\begin{figure*}[!ht]
    \centering
    \includegraphics[width=\textwidth]{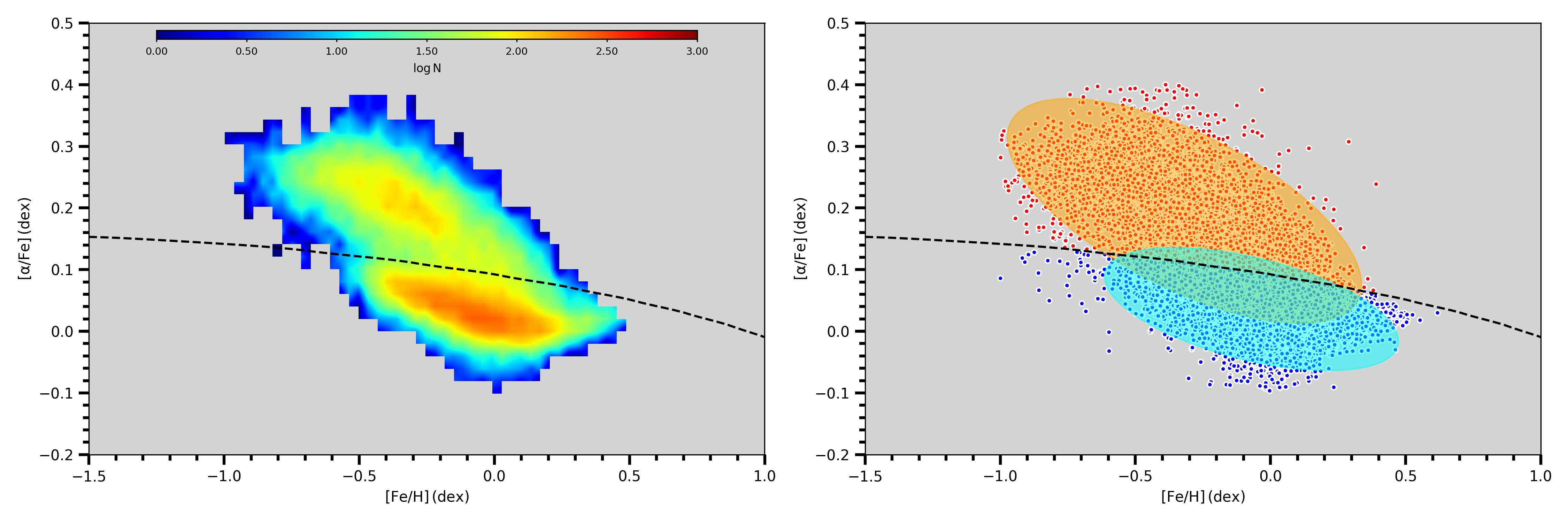}
    \caption{Distribution of RC stars on [$\alpha$/Fe]$\times$[Fe/H] plane. Left panel: Distribution is given for logarithmic number density. Black dashed line presents the decision boundary obtained with Gaussian mixture method. Right panel: Scatter diagram of RC stars is given with red and blue dots for high-$\alpha$ and low-$\alpha$ populations, respectively. Throughout the paper red and blue colours will represent these populations. Orange and blue ellipses represent the 1$\sigma$ probability distribution of each Gaussian for chemical space.}
    \label{fig:chemical_seperation}
\end{figure*}

This problem can be solved in our study by using the Gaussian Mixture Model (GMM), an unsupervised machine learning algorithm, to separate our RC sample into chemical populations. This algorithm classifies the data by fitting the desired number of Gaussian planes to the data. Classification is made by calculating the probabilities of the data of each Gaussian plane. Also, the different number of Gaussian sets can be fitted using different statistical methods. However, in this study, there is no need to do this because previous observational findings and observational models indicate that the Galaxy disc is composed of two separate chemical populations \citep{FBH02} that are mentioned in Section 3.1. In this study, a machine learning library {\it sklearn} version 0.19.1 \citep{scikit-learn} is used to apply GMM to the RC data. Model parameters are as follows: {\it n\_components} parameter, which specifies the number of entered classes, is set to 2; {\it covariance\_type}, which defines the covariance matrices of Gaussian surfaces while assigning each class, is selected as {\it Full}, because each plane is a separate covariance matrix and the planes intersect with each other. AGRC sample is separated into two classes as it is expected from the GMM and results are shown in Fig. \ref{fig:chemical_seperation}. In the figure, the orange and blue ellipses represent the $1\sigma$ probability distribution of the Gaussian planes determined by the model for two chemical populations, and the decision boundary is shown with a dashed black line. This boundary passes through the points where the possibilities of the two models are equal. As a result, GMM classified the region above this line as high-[$\alpha$/Fe] and below as low-[$\alpha$/Fe] (hereafter low-$\alpha$ and high-$\alpha$, respectively). According to GMM, there are 13,635 stars in low-$\alpha$, while 10,245 stars in high-$\alpha$ population. 

\section{Absolute Magnitudes and Colours from UV to MIR} 
In this study, stellar distances are estimated with $1/\varpi$ and BJ18 methods using {\it Gaia} DR2 trigonometric parallaxes. Absolute magnitudes of stars are obtained via Pogson's relation by using stellar distances, apparent magnitudes from UV to MIR, and interstellar extinction related coefficients from various authors (Table \ref{table:Table2}). Then, the median value of absolute magnitude distribution is determined for each chemical population in each photometric band. Building on these simple absolute magnitude estimations from the observational data, the RC distances are calculated by assuming a single absolute magnitude value for each chemical population where the results show deviation (i) as the population changes, (ii) as the photometric band changes, (iii) as the distance estimation method changes.

Since this study is aimed at investigating photometric distance determination for RC stars, it is important to perform an accurate estimate of  the interstellar extinction and reddening. Galactic dust maps give extreme values at low Galactic latitudes, even though various methods reduce these values. Stars above the Galactic plane, $|b|\geq10^\circ$, are selected in order to minimise the effect of interstellar extinction. The total extinction $A_{\infty}(b)$ in {\it V}-band for a star's direction is obtained from the dust map\footnote{https://irsa.ipac.caltech.edu/applications/DUST/} of  \citet{SF11} and estimated extinction $A_d(b)$ for the distance between Sun and the star using the equation of \citet{BS80} by adopting a scale-height of Galactic dust as $H$=125 pc \citep{Marshall06}. Details of the analysis used in estimating reduced extinction can be seen from \citet{Tuncel16}.

\begin{table}[!h]
\setlength{\tabcolsep}{4pt}
\centering
\caption{Coefficients of interstellar extinction for each photometric band from near UV to MIR wavelengths. Columns show photometric band, extinction coefficient and reference.}
\begin{tabular}{ccc}
\hline
Band & $A_{\lambda}/A_V$& Reference\\
\hline%
$NUV$& 2.335& \cite{Yuan13}\\
$u$  & 1.567& \cite{An09} \\
$g$  & 1.196&\cite{An09}  \\
$r$  & 0.874& \cite{An09} \\
$i$  & 0.672& \cite{An09} \\
$z$  & 0.488& \cite{An09} \\
$G$  & 0.859& \cite{Olivers19} \\
$J$  & 0.887& \cite{FM03} \\
$H$  & 0.565& \cite{FM03} \\
$K_s$& 0.382& \cite{FM03} \\
$W1$ & 0.039& \cite{Wang19} \\
$W2$ & 0.026& \cite{Wang19} \\
$W3$ & 0.040& \cite{Wang19} \\
\hline
\label{table:Table2}
\end{tabular}
\end{table}

\begin{figure*}[t]
    \centering
    \includegraphics[width=\textwidth]{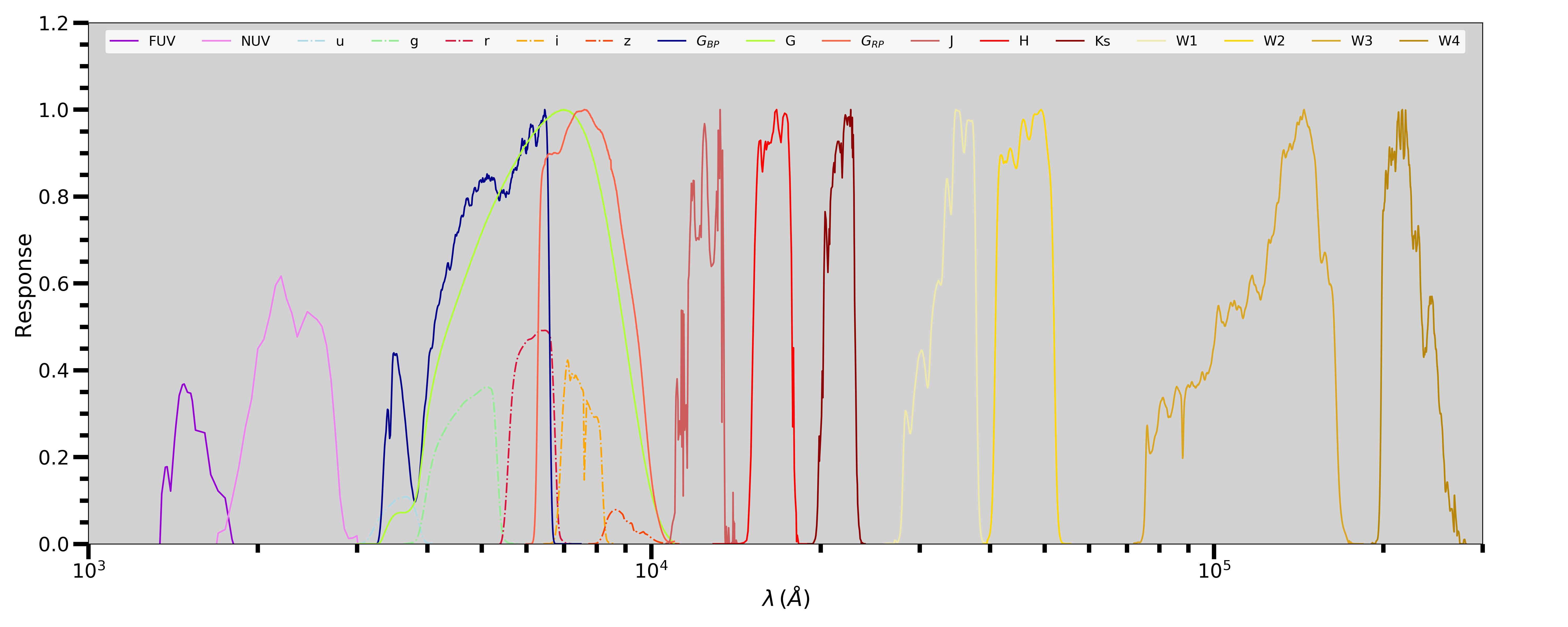}
    \caption{Response curves of photometric bands used in this study. Starting from left to right with {\it GALEX} $FUV$ (magenta) and $NUV$ (pink), SDSS $u$ (dashed light blue), $g$ (dashed green), $r$ (dashed red), $i$ (dashed yellow), $z$ (dashed orange) {\it Gaia} $G_{BP}$ (dark blue), $G$ (green), $G_{RP}$ (red), 2MASS $J$ (pale red), $H$ (bright red), $K_{S}$ (dark red) and {\it WISE} $W1$ (pale yellow), $W2$ (yellow), $W3$ (light brown), $W4$ (brown) bands.}
    \label{fig:band_response_curves}
\end{figure*}

Absolute magnitudes of RC stars are calculated using the conventional relation between apparent magnitude and distance in the form of $M_{\lambda} = m_{\lambda} - 5\log d + 5 - A_{\lambda}$. Here $M_{\lambda}$, $m_{\lambda}$, $d$ and $A_{\lambda}$ are absolute magnitude, apparent magnitude, distance and interstellar extinction in selected photometric band, respectively. Interstellar extinction coefficients for each photometric band are given in Table \ref{table:Table2}. Absolute magnitudes of RC stars in UV, optical, NIR and MIR of the electromagnetic spectrum are derived using data collected from {\it {\it GALEX}} GR6/7 \citep{Bianchi17}, SDSS DR7 \citep{Abazijan09}, {\it Gaia} DR2 \citep{GC18}, 2MASS \citep{Cutri03} and All-{\it WISE} \citep{Cutri13} surveys. Response curves of individual photometric bands of each photometric system are given in Fig. \ref{fig:band_response_curves}. In the study, the absolute magnitudes are determined by reading the median value of the band in question of the applied Gaussian fit on the frequency distribution in each individual wavelength range. Note that there are two main data sets, one is absolute magnitudes, and the other is colours. Each data set is further separated into low- and high-$\alpha$ populations with the method mentioned in Sect. 3.3, and then evaluated for the two distances that are obtained with different estimation methods. Overall results of median absolute magnitude and colour determinations are given in Table \ref{table:Table4} with consecutive panels. Each data set is evaluated specifically for the chosen surveys' unique conditions, i.e. quality flags, magnitude limits etc.; therefore sub-samples vary for each photometric band. Thus, based on this approach, the number of sources for each photometric system is different and unique, except for {\it WISE}. Number of stars vary with in each {\it WISE} bands.

\subsection{Ultraviolet Region}
The UV region is covered by the data of {\it GALEX} GR6/7 survey with far-ultraviolet ($FUV$, 170-300 nm) and  near-ultraviolet ($NUV$, 130-180 nm) photometric bands. Cross-match between AGRC sample and {\it GALEX} GR6/7 provides 8,899 RC stars with available observed magnitudes. Due to the lack of energy in $FUV$ band for many sources in the RC sample, $FUV-NUV$ colour could not be determined. Absolute magnitudes and colours are obtained for 8,716 sources in $NUV$ bands. RC stars are almost uniformly distributed in the  [$\alpha$/Fe]$\times$[Fe/H] plane, which allows a fair evaluation of the different parts of this chemically defined Galactic disc. There are about 4,200 stars in each chemical population for both distance sub-samples ($1/\varpi$ and BJ18). Absolute magnitude and colour distributions of {\it GALEX} $NUV$ sample for 1/$\varpi$ and BJ18 methods are given in Fig. \ref{fig:Fig8} and Fig. \ref{fig:Fig9}, respectively. Red coloured histograms represent the high-$\alpha$ population, while blue coloured histograms represent the low-$\alpha$ populations. Moreover, the Gaussian fit is shown with green solid line and the median value in each histogram is represented with turquoise dashed line. We will adopt the same notation also for the other absolute magnitude and colour distributions presented throughout the paper. The median distance of the low-$\alpha$ population is about 1.6 kpc, while for high-$\alpha$ population it is about 2.3 kpc, in 1/$\varpi$ method (Table \ref{table:Table4}). We find that stars in the sample have similar absolute magnitudes in $NUV$ for low- and high-$\alpha$ regimes in both distance sub-samples. Moreover, if each distance sub-sample is considered separately, the effect of the variation of chemical population causes 0.12 $\leq \Delta M \leq$ 0.13 mag variation in both distance method.

Even though no literature value to compare with $M_{NUV}$ magnitudes of this study is found except \citet[][]{Mohammed19}{'s} study. In their study, {\it GALEX} and {\it Gaia} photometric data are combined, and colour-metallicity relation is obtained for RC stars. However they results are incomparable with our study because they give only relation between $(NUV-G)_0$ and [Fe/H]. So, we made an inner comparison between 1/$\varpi$ and BJ18 methods for low-$\alpha$ and high-$\alpha$ magnitudes suggests that the UV peaks around the same median value with a very similar distribution for both chemical thin and thick disc populations. 

\subsection{Optical Region}
In this section, we present the results obtained using the SDSS DR7 and {\it Gaia} DR2 photometric bands. The Sloan photometric system covers $\lambda \lambda$ 300-1000 nm wavelength range with five photometric bands, i. e. $u$, $g$, $r$, $i$, $z$. Four photometric colours are defined from these bands, $(u-g)_o$, $(g-r)_o$, $(r-i)_o$ and $(i-z)_o$ \citep{Fukugita96, Fan99}. The cross-match between AGRC spectral catalogue and SDSS DR7 photometric catalogue returns 4,499 common RC stars. The distributions of SDSS absolute magnitudes are given in Fig. \ref{fig:Fig10} and Fig. \ref{fig:Fig11} for 1/$\varpi$ and BJ18 methods, respectively. Similarly, the distributions of SDSS colours are given in Fig. \ref{fig:Fig12} and Fig. \ref{fig:Fig13} for 1/$\varpi$ and BJ18 methods, respectively. There are almost two times more stars in low-$\alpha$ than high-$\alpha$ population. For the distances estimated by the 1/$\varpi$ and BJ18 method, absolute magnitudes in SDSS $u$ band appear distinct in behaviour than the rest of the remaining bands. The lower sensitivity of the $u$-band does not allow a more reliable data evaluation. Our results indicate that the high-$\alpha$ population is brighter than the low-$\alpha$ one in the $u$-band. In $g$ band, RC stars become brighter by 0.09 mag as the chemical population changes from low- to high-$\alpha$. A similar trend is followed by $r$, $i$ and $z$ bands, with an increase in brightness by 0.04, 0.05 and 0.10 mag, respectively. This relatively larger absolute magnitude difference value in the $z$ band might be caused by low response of this photometric band, which is shown in Fig. \ref{fig:band_response_curves} with orange dashed line.

\begin{table*}[!ht]
\setlength{\tabcolsep}{1pt}
{\scriptsize
\centering
\caption{{\it Top-Left panel}: Absolute magnitudes obtained for low-$\alpha$ and high-$\alpha$ populations for {\it GALEX} GR6/7, SDSS DR7,  {\it Gaia} DR2, 2MASS, and All-{\it WISE} photometric surveys using 1/$\varpi$ distances. {\it Top-Right panel}: Same with top-left panel except it shows the absolute magnitude values for BJ18 distances. {\it Bottom-Left panel}: Colours obtained for low-$\alpha$ and high-$\alpha$ populations for {\it GALEX} GR6/7, SDSS DR7, {\it Gaia} DR2, 2MASS, and All-{\it WISE} photometric surveys using 1/$\varpi$ distances. {\it Bottom-Right panel:} Same with top-left panel except it shows the colour values for BJ18 distances. First columns give the survey name and second columns give absolute magnitudes (top panels) or defined colours (bottom panels) for selected photometric system in all four panels. Then, each panel is divided into low-$\alpha$ and high-$\alpha$ populations. For each $\alpha$ population, median absolute magnitude or median colour, it's standard deviation ($\sigma$), standard error (SE), number of stars ($N$), median distance ($d$), difference between high-$\alpha$ and low-$\alpha$ absolute magnitudes ($\Delta M$) and colours ($\Delta C$) are listed, respectively.}

\begin{tabular}{c|c|ccccc|ccccc|c||ccccc|ccccc|c}
\cline{3-24}
\cline{3-24}
\multicolumn{2}{c}{}&\multicolumn{11}{c||}{1/$\varpi$}&\multicolumn{11}{c}{BJ18} \\
\cline{3-24}
\cline{3-24}
\multicolumn{2}{c}{}&\multicolumn{5}{c}{low-$\alpha$}& &\multicolumn{5}{c||}{high-$\alpha$}&\multicolumn{5}{c}{low-$\alpha$} &\multicolumn{5}{c}{high-$\alpha$} &\\
\hline
 Survey & Absolute & $\tilde{M}$ & $\pm \sigma$ & SE & $N$ & $\tilde{d}$ & $\tilde{M}$ & $\pm \sigma$ & SE & $N$ & $\tilde{d}$ & $\Delta M$ & $\tilde{M}$ & $\pm \sigma$ & SE & $N$ & $\tilde{d}$ & $\tilde{M}$ & $\pm \sigma$ & SE& $N$ & $\tilde{d}$ & $\Delta M$ \\
        & magnitude & mag & mag & mag & stars & pc & mag & mag & mag & stars & pc & mag & mag & mag & mag & stars & pc & mag & mag & mag & stars & pc & mag \\
\hline
 {\it GALEX} & $M_{NUV}$  &  8.49 & 0.88 & 0.013 & 4400 & 1632 & 8.37 & 0.88 & 0.013 & 4316 & 2298 & 0.12 & 8.50 & 0.88 & 0.073 & 4231 & 1548 & 8.38 & 0.88 & 0.166 & 4137 & 2140 & 0.13\\
SDSS & $M_u$  & 4.70 & 0.84 & 0.015  & 2951 & 1168 &  4.02 & 0.82  & 0.024 & 1548 & 1793 &  0.68 &  4.78 & 0.85  & 0.025  & 1130 & 1164 & 4.24  & 0.80  & 0.032 & 615 & 1653 &0.54 \\
     & $M_g$  & 1.45 & 0.39 & 0.007  & 2951 & 1168 &  1.54 & 0.51  & 0.015 & 1548 & 1793 &  -0.09 &  1.53 & 0.41  & 0.012  & 1130 & 1164 & 1.62  & 0.44  & 0.018 & 615 & 1653 & -0.09 \\
     & $M_r$  & 0.48 & 0.26 & 0.005  & 2951 & 1168 &  0.52 & 0.32  & 0.009 & 1548 & 1793 &  -0.04 &  0.55 & 0.25  & 0.007  & 1130 & 1164 & 0.64  & 0.24  & 0.010 & 615 & 1653 & -0.09 \\
     & $M_i$  & 0.20 & 0.24 & 0.004  & 2951 & 1168 &  0.25 & 0.30  & 0.009 & 1548 & 1793 &  -0.05 &  0.28 & 0.24  & 0.007  & 1130 & 1164 & 0.37  & 0.22  & 0.009 & 615 & 1653 & -0.09 \\
     & $M_z$  & 0.58 & 0.34 & 0.006  & 2951 & 1168 &  0.68 & 0.43  & 0.010 & 1548 & 1793 &  -0.10 &  0.65 & 0.37  & 0.011  & 1130 & 1164 & 0.79  & 0.39  & 0.016 & 615 & 1653 & -0.14 \\
{\it Gaia} & $M_G$  & 0.45 & 0.18 & 0.002  &11290 & 1783 &  0.51 & 0.15 & 0.004 & 6896 & 2391 & -0.06 & 0.54 & 0.21 & 0.002 & 13432 & 1738 & 0.64  & 0.19  & 0.002 & 10114 & 2243 & -0.10\\
 2MASS     & $M_J$  & -1.17 & 0.24 & 0.002 & 12550 & 1845 & -1.05 & 0.30 & 0.003 & 9624 & 2405 & -0.12 & -1.05 & 0.21 & 0.002   & 12368 & 1751 & -0.89  & 0.27 & 0.003& 9506 &  2244 & -0.16 \\
           & $M_H$  & -1.68 & 0.23 & 0.002 & 12550 & 1845 & -1.56 & 0.30 & 0.003 & 9624 & 2405 & -0.12 & -1.56 & 0.21 & 0.002.  & 12368 & 1751  & -1.40  & 0.27 & 0.003& 9506 & 2244  & -0.16 \\
          &$M_{K_s}$& -1.79 & 0.22 & 0.002 & 12550 & 1845 & -1.65 & 0.30 & 0.003 & 9624 & 2405 & -0.14 &  -1.67 & 0.20 & 0.002  & 12368 & 1751 & -1.50  & 0.28 & 0.003& 9506 & 2244 & -0.17 \\
{\it WISE}      & $M_{W1}$& -1.84 & 0.22 & 0.002 & 13327 & 1632 & -1.71 & 0.30 & 0.003 & 10146 & 2298& -0.13 &-1.72 & 0.19 & 0.002  & 12256 & 1731 & 1.55 & 0.28 &0.003 &9141 & 2200 & -0.17 \\
          & $M_{W2}$& -1.74 & 0.22 & 0.002 & 13438 & 1632 & -1.61  & 0.29 & 0.003 & 10228 & 2298 & -0.13 &-1.62 & 0.19 & 0.002  & 12325 & 1731 & -1.45 & 0.27 &0.003 & 9214 & 2200 & 0.17 \\
          & $M_{W3}$& -1.85 & 0.22 & 0.002 & 12840 & 1826 & -1.74  & 0.30 & 0.003 &  8570 & 2405& -0.11 &-1.74 & 0.20 & 0.002  & 11696 & 1682 & -1.59 & 0.28 &0.003 &7903 & 2066 & 0.15 \\
\hline
\hline
\multicolumn{2}{c}{}&\multicolumn{5}{c}{low-$\alpha$}&&\multicolumn{5}{c||}{high-$\alpha$}&\multicolumn{5}{c}{low-$\alpha$} &\multicolumn{5}{c}{high-$\alpha$} &\\
\hline
 Survey & Colour & $\tilde{C}$ & $\pm \sigma$ & SE &$N$ && $\tilde{C}$ & $\pm \sigma$ & SE & $N$ && $\Delta C$ & $\tilde{C}$ & $\pm \sigma$ & SE & $N$  && $\tilde{C}$ & $\pm \sigma$ & SE & $N$ && $\Delta C$ \\
        & index & mag & mag & mag & stars &  & mag & mag & mag & stars &  & mag & mag & mag & mag & stars &  & mag & mag & mag & stars &  & mag \\
\hline
 {\it GALEX}    & ---  & --- & --- & --- & --- & --- & --- & --- & ---& --- & --- & --- & --- &---& --- & --- & --- & --- & --- & --- & --- & --- & ---\\
\hline
SDSS & $(u-g)_o$ & 3.32 & 0.83 & 0.015 & 2951 && 2.57 & 0.98 & 0.029 & 1548 &&  0.75 & 3.37 & 0.88  & 0.026  & 1130 &&  2.68  & 0.89  & 0.036 &615& & 0.69\\
     & $(g-r)_o$ & 0.92 & 0.18 & 0.003 & 2951 && 0.87 & 0.16 & 0.005 & 1548 &&  0.05 & 0.93 & 0.18  & 0.005  & 1130 &&  0.87  & 0.16  & 0.006 &615 && 0.06\\
     & $(r-i)_o$ & 0.28 & 0.03 & 0.001 & 2951 && 0.28 & 0.03 & 0.001 & 1548 &&  0.00 &  0.28 & 0.03  & 0.001  & 1130 &&  0.27  & 0.03  & 0.001 &615 && 0.01\\
     & $(i-z)_o$ &-0.34 & 0.25 & 0.005 & 2951 &&-0.37 & 0.28 & 0.008 & 1548 &&  0.03 &-0.32 & 0.24  & 0.007  & 1130& & -0.35  & 0.27  & 0.011 &615& & 0.03\\
{\it Gaia} & $(G_{BP}-G_{RP})_o$ & 1.22 & 0.04 & 0.001& 11290 && 1.21 & 0.04 & 0.001& 6896&&  0.01 & 1.22 & 0.04  & 0.001 & 13432&&  1.20 & 0.06 & 0.001 & 10114&& 0.02\\

 2MASS & $(J-K_s)_o$ & 0.62 & 0.05 & 0.001 & 12550 && 0.62 & 0.05 & 0.001 & 9624 &&  0.00    & 0.62 & 0.05 & 0.001 & 12368 &&  0.62 & 0.05 & 0.001 & 9506 &&   0.00   \\
       & $(H-K_s)_o$ & 0.10 & 0.03 & 0.001 & 12550 && 0.10 & 0.03 & 0.001 & 9624 &&  0.00    & 0.10 & 0.03 & 0.001 & 12368& &  0.10 & 0.03 & 0.001 & 9506 &&   0.00 \\
       & $(J-H)_o$   & 0.51 & 0.04 & 0.001 & 12550 && 0.52 & 0.04 & 0.001 & 9624 && -0.01 & 0.51 & 0.04 & 0.001 & 12368 &&  0.52 & 0.04 & 0.001 & 9506 &&  -0.01\\
{\it WISE}       & $(W1-W2)_o$   & -0.10 & 0.02  & 0.001 & 13327 && -0.09 & 0.02 & 0.001 & 10146 && -0.01&  -0.10 &0.02 & 0.001 & 13309 &&  -0.09&0.02  & 0.001 &10126 & & -0.01\\
           & $(W1-W3)_o$   &  0.01 & 0.04  & 0.001 & 12539 &&  0.03 & 0.04 & 0.001 & 8478  && -0.02&   0.01 &0.04 & 0.001 & 12526 &&   0.03&0.04  &  0.001& 8469 & & -0.02\\
           & $(W2-W3)_o$   &  0.11 & 0.03  & 0.001 & 12643 &&  0.12 & 0.04 & 0.001 & 8552  && -0.01&   0.11 &0.03 & 0.001 & 12629 &&   0.12&0.04  & 0.001 & 8543 & &  -0.01\\
\hline
\hline
\label{table:Table4}
\end{tabular}
}
\end{table*}

No literature value is found for $u$-band to compare as can be seen from Table \ref{table:Table7}, but the inner comparison between 1/$\varpi$ and BJ18 methods for the chemical thin disc suggest that there is a 0.08 mag difference, while for the chemical thick disc is even larger, up to 0.22 mag. SDSS $M_g$, $M_r$, $M_i$ and $M_z$ magnitudes are compared with the results of \citet{Chen17} with 171 RC stars with APASS Sloan photometry, and the result of \citet{Ruiz18}, with 1,315 TGAS DR1 RC stars. These studies are focused on the thin disc sample, so that their results are useful only for the low-$\alpha$ populations. $M_g$ magnitudes of \citet{Chen17} and \citet{Ruiz18} are brighter about 0.20 mag than our low-$\alpha$ RC's for 1/$\varpi$ method. These values are even larger (by 0.08 mag)  when we consider the results obtained with BJ18 method. Absolute magnitudes in $r$ and $i$ bands coincide with the chemical thin disc values obtained with BJ18's method. On the other hand, the absolute magnitudes in $z$ band are quite different ($\Delta M_z > 0.60$ mag) from the \citet{Chen17} result. 

Absolute magnitudes and colours in optical bands are further analysed using the {\it Gaia} $G$-band magnitude and $(G_{BP}-G_{RP})_o$ colour, which cover the same wavelength range of SDSS, but with much broader photometric bands. After the chemical populations are separated, 62\% of {\it Gaia} sample stars are in low-$\alpha$ while 38\% is in high-$\alpha$ population, which is expected due to the astrometric precision of this RC sample. The distributions for $G$-band absolute magnitude and $(G_{BP}-G_{RP})_o$ de-reddened colour are shown in Fig. \ref{fig:Fig14} and Fig. \ref{fig:Fig15}, respectively. The $G$-band absolute magnitudes increase by 0.06 mag for the 1/$\varpi$ method while this difference is 0.1 mag for BJ18 method, from low-$\alpha$ to high-$\alpha$ population (Table \ref{table:Table4}). On the other hand, $(G_{BP}-G_{RP})_o$ colours appear to be not affected by either the chemical population or the distance estimation method. As seen Table \ref{table:Table4}, the resulting colour difference is almost constant for both methods.

In {\it Gaia} $G$ band, the absolute magnitude for low-$\alpha$ RC stars for 1/$\varpi$ method seems to be in good agreement with the \citet{Hawkins17} and \citet{Ruiz18}'s results that were obtained using TGAS catalogue \citep{Michalik15}. Optic and UV absolute magnitudes, and literature comparison is given in Table \ref{table:Table7}.

\subsection{Infrared Region}
\begin{figure*}[!t]
    \centering
    \includegraphics[width=\textwidth]{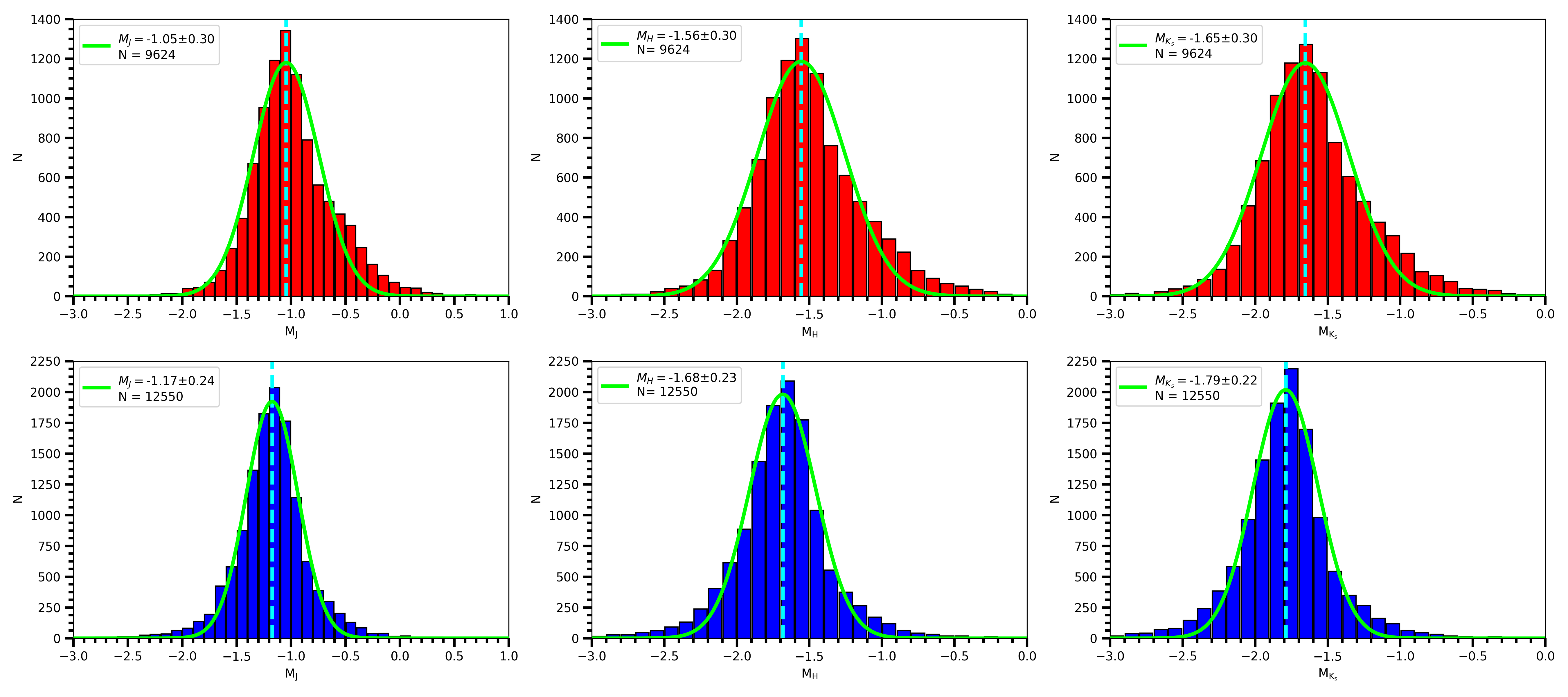}
\caption{Absolute magnitude distribution of RC stars in 2MASS $J$, $H$, and $K$ bands that are calculated with the 1/$\varpi$ method for high-$\alpha$ (upper panel) and low-$\alpha$ (lower panel) populations. Green solid line is the Gaussian fit for the distribution and turquoise dashed line is the median value of the distribution.} 
\label{fig:Fig16}
\end {figure*}

\begin{figure*}[!t]
    \centering
    \includegraphics[width=\textwidth]{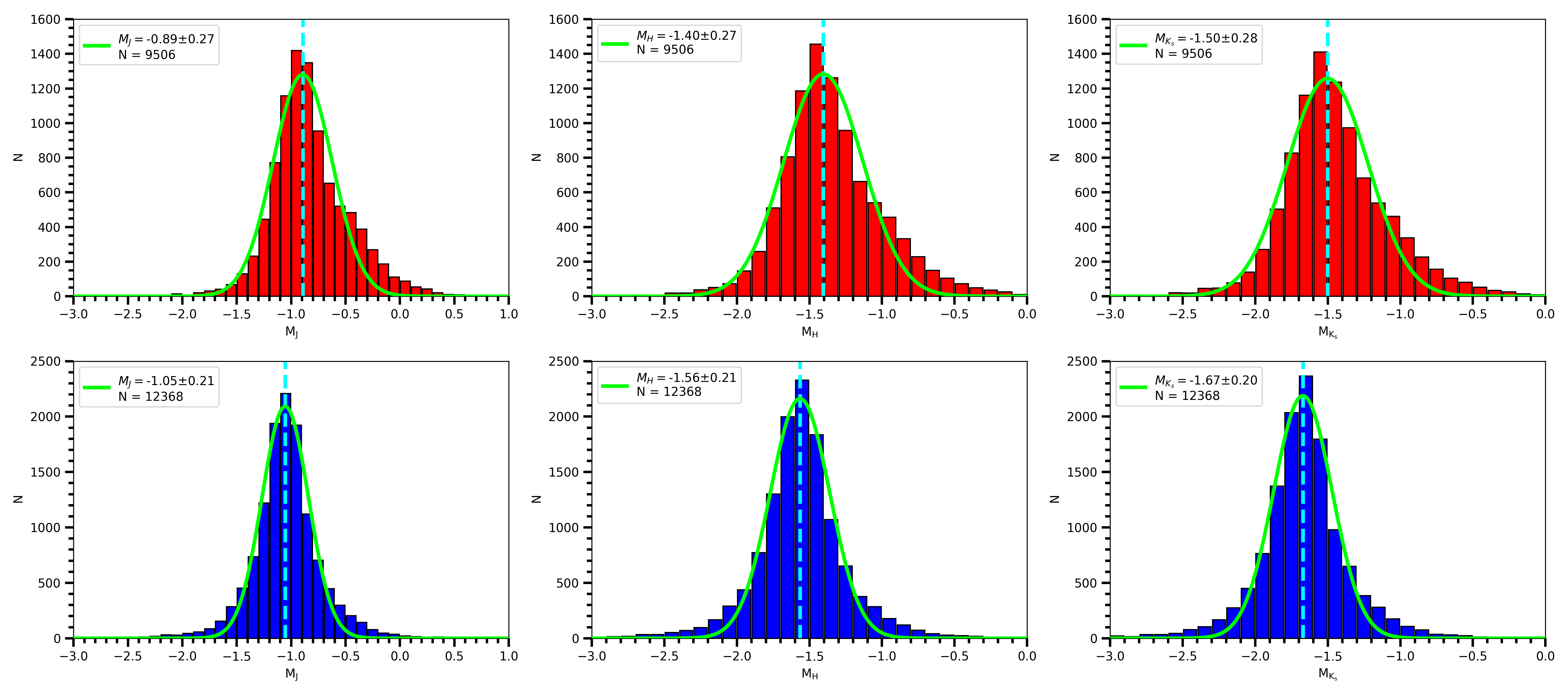}
\caption{Absolute magnitude distribution of RC stars in 2MASS $J$, $H$, and $K$ bands that are calculated with the BJ18 method for high-$\alpha$ (upper panel) and low-$\alpha$ (lower panel) populations. Green solid line is the Gaussian fit for the distribution and turquoise dashed line is the median value of the distribution.} 
\label{fig:Fig17}
\end {figure*}

For the analysis of the NIR region we use  the 2MASS photometric bands $J$ (1.25 $\mu$m), $H$ (1.65 $\mu$m) and $K_s$ (2.17 $\mu$m). Stars with photometric quality  AAA are selected for their more accurate photometry \citep{Cutri03}. More than 22,000 RC stars are cross-matched with AGRC spectral catalogue and they are almost equally distributed in the chemical space. Three absolute magnitudes, i. e. $M_J$, $M_H$ and $M_{K_s}$ and three  de-reddened colour indices, i.e. $(J-K_{s})_o$, $(J-H)_o$ and $(H-K_{S})_o$ are calculated with the method described in Sect. 4. For 2MASS photometry, most of the RC sample have all three apparent magnitudes. Thus there is a unique number of stars for each chemical population under each distance estimation method. The low-$\alpha$ population contains more than 12,500 RC stars while high-$\alpha$ population has about 9,600 RC stars. Based on the absolute magnitude frequency distributions, which are given in Figs. \ref{fig:Fig16} and \ref{fig:Fig17}, chemical thin disc RC stars are brighter by 0.12 or 0.14 mag than the chemical thick disc stars once the distances are estimated via 1/$\varpi$ method. On the other hand, this value is pushed to 0.16 or 0.17 mag difference when using the BJ18 method. Based on the colour distributions given in Figs. \ref{fig:Fig18} and \ref{fig:Fig19}, there is no clear variation in 2MASS colours (Table \ref{table:Table3}).

\begin{table*}[!ht]
\setlength{\tabcolsep}{5pt}
\scriptsize
\centering
\caption{Literature comparison of absolute magnitudes in near-UV and optical bands obtained from low-$\alpha$ and high-$\alpha$ populations for 1/$\varpi$ and BJ18 methods. Columns give survey, number of stars, absolute magnitudes and remarks that include the data source.}
\begin{tabular}{lccccccr}
\hline
\multicolumn{1}{c}{\it GALEX}                   & $N$ (stars) &  $M_{NUV}$ (mag) &  &  &  &  & \multicolumn{1}{c}{Remarks} \\
\hline
This study low-$\alpha$            & 1632 &  8.49$\pm$0.013  &  &  &  &  & {\it Gaia} DR2 (1/$\varpi$) \\
This study high-$\alpha$           & 2298 &  8.37$\pm$0.013  &  &  &  &  & {\it Gaia} DR2 (1/$\varpi$) \\
This study low-$\alpha$            & 1548 &  8.50$\pm$0.073  &  &  &  &  & {\it Gaia} DR2 (BJ18) \\
This study high-$\alpha$           & 2140 &  8.38$\pm$0.166  &  &  &  &  & {\it Gaia} DR2 (BJ18) \\
\hline
\multicolumn{1}{c}{SDSS}                    & $N$ (stars)  &  $M_{u}$ (mag)      & $M_{g}$ (mag)       & $M_{r}$ (mag)       & $M_{i}$ (mag)    & $M_{z}$ (mag)        & \multicolumn{1}{c}{Remarks}\\
\hline 
\citet{Chen17}          & 171  & ---            & 1.229$\pm$0.172 & 0.420$\pm$0.110 & 0.157$\pm$0.094 & 0.022$\pm$0.084 & 171 APASS $ugriz$ \\
\citet{Ruiz18}          & 1315 & ---            & 1.331$\pm$0.056 & 0.552$\pm$0.026 & 0.262$\pm$0.032 & ---             & TGAS {\it Gaia}   DR1\\
This study low-$\alpha$            & 1168 & 4.70$\pm$0.015  & 1.45$\pm$0.007   & 0.48$\pm$0.005   & 0.20$\pm$0.004   & 0.58$\pm$0.006 & {\it Gaia} DR2 (1/$\varpi$) \\
This study high-$\alpha$           & 1793 & 4.02$\pm$0.024  & 1.54$\pm$0.015   & 0.52$\pm$0.009   & 0.25$\pm$0.009   & 0.68$\pm$0.010 & {\it Gaia} DR2 (1/$\varpi$) \\
This study low-$\alpha$            & 1164 & 4.78$\pm$0.025  & 1.53$\pm$0.012   & 0.55$\pm$0.007   & 0.28$\pm$0.007   & 0.65$\pm$0.011 & {\it Gaia} DR2 (BJ18) \\
This study high-$\alpha$           & 1653 & 4.24$\pm$0.032  & 1.62$\pm$0.018   & 0.64$\pm$0.010   & 0.37$\pm$0.009   & 0.79$\pm$0.016 & {\it Gaia} DR2 (BJ18) \\
\hline
\multicolumn{1}{c}{\it Gaia}      & $N$ (stars)  & $M_G$ (mag) & & & & & \multicolumn{1}{c}{Remarks} \\
\hline
\citet{Hawkins17}       & 972    &	0.44$\pm$0.01  & & & & &  TGAS   \\
\citet{Ruiz18}          & 2482   & 0.495$\pm$0.009 & & & & &  TGAS    \\
This study low-$\alpha$            & 11290  & 0.45$\pm$0.002   & & & & &  {\it Gaia} DR2 (1/$\varpi$) \\
This study high-$\alpha$           & 6896   & 0.51$\pm$0.004   & & & & &  {\it Gaia} DR2 (1/$\varpi$)  \\
This study low-$\alpha$            & 13432  & 0.54$\pm$0.002   & & & & &  {\it Gaia} DR2 (BJ18)   \\
This study high-$\alpha$           & 10114  & 0.64$\pm$0.002   & & & & &  {\it Gaia} DR2 (BJ18)  \\

\hline
\hline
\label{table:Table7}
\end{tabular}
\end{table*}

\begin{table*}[!ht]
\setlength{\tabcolsep}{3pt}
\scriptsize
\centering
\caption{Literature comparison of absolute magnitudes in NIR and MIR bands obtained from low-$\alpha$ and high-$\alpha$ populations for 1/$\varpi$ and BJ18 methods. Columns give survey, number of stars, absolute magnitudes and remarks that include the data source.}
\begin{tabular}{lccccccr}
\hline
\multicolumn{1}{c}{2MASS}                 & $N$ (stars)       & $M_{J}$ (mag)         & $M_{H}$ (mag)           &   $M_{K_s}$ (mag)      & & & \multicolumn{1}{c}{Remarks}\\
\hline
\citet{Alves00}       & 238     & 	-                &   -               & -1.61$\pm$0.03   & & & {\it Hipparcos} \\
\citet{GS02}          & 14	    &   -      	         &   -               & -1.61$\pm$0.04   & & &  WYIN open clusters\\
\citet{vHG07}         & 24      &   -                &   -               & -1.57$\pm$0.05   & & & 2MASS open clusters\\
\citet{Cabrera07b}     & 49*    &	-                &   -               & -1.62$\pm$0.03   & & & TCS-CAIN\\
\citet{Groenewegen08} &         &	-                &   -               & -1.54$\pm$0.04   & & & revised {\it Hipparcos}\\
\citet{Laney12}       & 191     &	-0.984$\pm$0.014 &	-1.490$\pm$0.015  & -1.613$\pm$0.015 & & & revised {\it Hipparcos}\\
\citet{YG13}          & 2937    & -0.970$\pm$0.016   &	-1.462$\pm$0.014 & -1.595$\pm$0.025 & & & revised {\it Hipparcos}\\
\citet{FA14}          &         &	-                &   -               & -1.53$\pm$0.01   & & & revised {\it Hipparcos}\\
\citet{Chen17}        & 171	    & -1.016$\pm$0.063   &	-1.528$\pm$0.055 & -1.626$\pm$0.057 & & & SAGA\\
\citet{Hawkins17}     & 972     & -0.93$\pm$0.01     &	-1.46$\pm$0.01   & -1.61$\pm$0.01   & & &  TGAS  \\
\citet{Ruiz18}        & 2482	& -0.945$\pm$0.01    &	-1.450$\pm$0.017 & -1.606$\pm$0.009 & & &  TGAS \\
This study low-$\alpha$         & 12550   & -1.17$\pm$0.002     &  -1.68$\pm$0.002   & -1.79$\pm$0.002   & & &  {\it Gaia}   DR2 (1/$\varpi$) \\
This study high-$\alpha$        & 9624    & -1.05$\pm$0.003     &  -1.56$\pm$0.003   & -1.65$\pm$0.003   & & &  {\it Gaia}   DR2 (1/$\varpi$)\\
This study low-$\alpha$         & 12638   & -1.05$\pm$0.002     &  -1.56$\pm$0.002   & -1.67$\pm$0.002   & & & {\it Gaia}   DR2 (BJ18)   \\
This study high-$\alpha$        &  9506   & -0.89$\pm$0.003     &  -1.40$\pm$0.003   & -1.50$\pm$0.003   & & & {\it Gaia}   DR2 (BJ18)   \\
\hline
\multicolumn{1}{c}{\it WISE}                  & $N$ (stars)       & $M_{W1}$ (mag)           & $M_{W2}$ (mag)         & $M_{W3}$ (mag)         & $M_{W4}$ (mag)         &  & \multicolumn{1}{c}{Remarks} \\
\hline
\citet{YG13}          & 3889    & -1.612$\pm$0.022   &	-                &	-1.585$\pm$0.019 &	-                &  &  revised {\it Hipparcos} \\
\citet{Chen17}        & $<$171    & -1.694$\pm$0.061   &	-1.595$\pm$0.064 &	-1.752$\pm$0.068 &	-                &  &  SAGA \\
\citet{Hawkins17}     & $>$900    & -1.68$\pm$0.02     &  -1.69$\pm$0.02	 &  -1.67$\pm$0.01	 &  -1.76$\pm$0.01   &  &  TGAS  \\
\citet{Ruiz18}        & 962, 1032, 2026, 747& -1.711$\pm$0.017 &  -1.585$\pm$0.016   &	-1.638$\pm$0.011 &	-1.704$\pm$0.012 &  & TGAS \\
This study low-$\alpha$          & [12840, 13438]   & -1.84$\pm$0.002     &  -1.74$\pm$0.002   &  -1.85$\pm$0.002   &  -  &  & {\it Gaia} DR2 (1/$\varpi$) \\
This study high-$\alpha$         & [8570, 10228]   & -1.71$\pm$0.003     &  -1.61$\pm$0.003   &  -1.74$\pm$0.003   &  -  &  & {\it Gaia} DR2 (1/$\varpi$) \\
This study low-$\alpha$          & [11696, 12325]   & -1.72$\pm$0.002     &  -1.62$\pm$0.002   &  -1.74$\pm$0.002   &  -   &  &  {\it Gaia} DR2 (BJ18)  \\
This study high-$\alpha$         & [7903, 9214]    & -1.55$\pm$0.003     &  -1.45$\pm$0.003   &  -1.59$\pm$0.003   &  -   &  &   {\it Gaia}   DR2 (BJ18)    \\
\hline 
\hline
\label{table:Table8}
\end{tabular}
\end{table*}

Literature values are compiled in Table \ref{table:Table8} for NIR and MIR regions. The standard candle status of RC stars is extensively studied with 2MASS photometry since the beginning of the new millennium. Early studies used the astrometric data provided by {\it Hipparcos} survey, TGAS or open cluster data. According to the literature studies, $M_J$ values vary between -0.93 to -1.02 mag for the thin disc \citep{Laney12, Bilir13b, Karaali13, YG13, Chen17, Hawkins17, Ruiz18}. The median $M_J$ value for the chemical thin disc is -1.17 mag (for 1/$\varpi$), which is 0.15 mag brighter than that in the literature while, for the BJ18 method, it is -1.05 mag, which is closer to the literature values. Absolute magnitudes in $H$-band vary between -1.47 and -1.53 mag (see Table \ref{table:Table8}). The $M_H$ absolute magnitudes of low-$\alpha$ sub-sample (for 1/$\varpi$) are cover 0.15 and 0.23 mag brighter than the existing values in literature, respectively. On the other hand, low-$\alpha$ sub-sample (for BJ18) is close to the values of \citet{Chen17}'s RC stars that were selected with asteroseismology methods. $M_{K_s}$ is the most studied magnitude to test the standard candle status of RC stars. The majority of literature studies agree on $M_{K_s} = -1.61$ mag value \citep{Alves00, GS02, Laney12, YG13, Chen17, Hawkins17, Ruiz18} while, other studies give brighter values but within the error bars. However, the results for $M_{K_s}$ absolute magnitude obtained in this study are -1.79 and -1.65 mag for chemical thin disc and -1.67 and -1.50 mag for chemical thick disc. These values are not in agreement with any literature values.

For the analysis of the MIR region we use  the four WISE photometric bands, $W1$ (3.368 $\mu$m), $W2$ (4.618 $\mu$m), $W3$ (12.082 $\mu$m), and $W4$ (22.194 $\mu$m) and three colours, i. e. $W1-W2$, $W1-W3$, and $W2-W3$. The {\it WISE} sample is selected to obtain an optimum number of stars in each photometric band. Overall, there are more objects in the low-$\alpha$ population ($\approx12,500$) than high-$\alpha$ ($\approx9,200$) for both distance method in {\it WISE} photometry. However, for $W4$ photometric band, the number of RC stars are not sufficient to perform a statistically meaningful analysis. Frequency distribution of absolute magnitudes in $W1$, $W2$, and $W3$ photometric bands are shown in Figs. \ref{fig:Fig20} and \ref{fig:Fig21}.

Absolute magnitudes vary -0.11 and -0.13 mag for {\it WISE} bands as the population changes from low- to high-$\alpha$ for 1/$\varpi$ method. This difference increases between -0.15 and -0.17 mag using the BJ18 method. Regardless of the number of RC stars, colours are changed mildly ($< 0.02$ mag) from low- to high-$\alpha$ population, as shown in Figs. \ref{fig:Fig22} and \ref{fig:Fig23}.


{\it WISE} absolute magnitudes are obtained from the samples with {\it Hipparcos} trigonometric parallaxes and with relatively small RC samples (less than 4,000), Three of the four studies obtain the distances using photometric parallax, inverse parallax, and InfraRed flux methods. Based on the literature results in Table \ref{table:Table8}, for the $M_{W1}$ values by \citet{Chen17}, \citet{Hawkins17}, \citet{Ruiz18} results scatter around -1.69 mag, while \citet{YG13}'s $M_{W1}$ is -1.61 mag. For the BJ18 method, the value for the chemical thin disc sample is in agreement literature, but it is 0.10 mag brighter for the chemical thin disc of $1/\varpi$ method. In $M_{W2}$, chemical thin disc of BJ18 is in agreement with \citet{Chen17} and \citet{Ruiz18}, while chemical thin disc of $1/\varpi$ is in accordance with \citet{Hawkins17}'s studies.

Overall comparison shows that regardless of the distance method, we found a general trend of increasing luminosity from shorter to longer  wavelengths, with a relative magnitude variation between chemical population.



\section{Testing Absolute Magnitudes with a Mock Catalogue}
 The mock data is generated with the {\it Besan\c con} population synthesis model \citep{Robin03,Robin12}. This model provides theoretical density distribution for stars in  the Galactic bulge, bar, thin disc, thick disc, and halo. The most up to date version of the model is run for the most crowded Galactic region in our AGRC catalogue, $90^\circ \leq l \leq 270^\circ$. This model is chosen because it can produce [$\alpha$/Fe] with steller atmospheric parameters. These parameters allow us to apply the same procedure to select synthetic RC stars that belong to different chemical populations from the mock sample like our AGRC catalogue.

\begin{figure*}[t]
    \centering
    \includegraphics[scale=0.4]{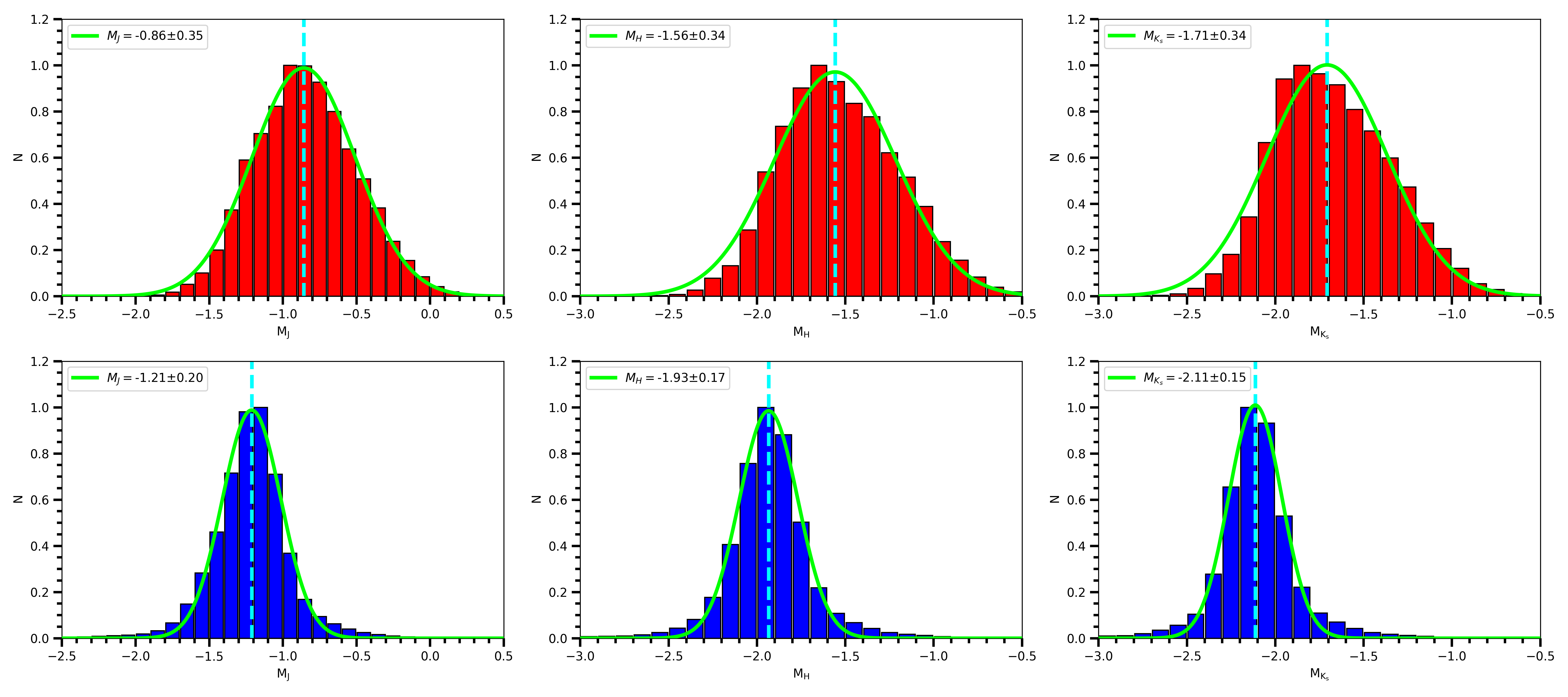}
\caption{Absolute magnitude distributions of RC stars selected from {\it Besan\c con} population synthesis model for 2MASS photometric system for high-$\alpha$ (upper panel) and low-$\alpha$ (lower panel) populations.} 
\label{fig:besancon_rc}
\end {figure*}


2MASS photometry, with its high $S/N$ as a base photometric system for APOGEE and GALAH spectroscopic surveys, is chosen to compare the model with observational data. Model results are plotted on the HR diagram, and RC stars are selected with the same method used in Sect. 2.1. Chemical population separation is done by applying the same criteria used in Sect. 3 to the mock catalogue. In Fig. \ref{fig:besancon_rc}, absolute magnitudes for 2MASS photometry of the mock catalogue are shown for low- and high-$\alpha$ RC populations. Results of this mock sample indicate that the observed absolute magnitude differences in both populations still hold like in the observational data. However, the difference is more amplified, i.e. $\langle \Delta M \rangle = -0.37$ mag. Comparison between the observational and theoretical absolute magnitudes in 2MASS bands is given in Table \ref{table:Table3}.

\begin{table}[h]
\setlength{\tabcolsep}{4pt}
\centering
\caption{Comparison between 2MASS $M_J$, $M_H$, and $M_{K_s}$ absolute magnitudes generated with {\it Besan\c con} population synthesis model and AGRC catalogue for low-$\alpha$ and high-$\alpha$ populations. Absolute magnitude errors are less than 0.003 mag.}
\begin{tabular}{l|cc|cc|cc|} 
\cline{2-7}
        & \multicolumn{2}{c|}{{\it Gaia}} & \multicolumn{2}{c|}{{\it Besan\c con}} & \multicolumn{2}{c|}{$\Delta M$ (mag)} \\
\cline{2-7}
        & low-$\alpha$ & high-$\alpha$ & low-$\alpha$ & high-$\alpha$ & observed & model  \\
\cline{1-7}
 $M_J$     & -1.17 & -1.05 &  -1.21 & -0.86 & -0.12 & -0.35\\
 $M_H$     & -1.68 & -1.56 &  -1.93 & -1.56 & -0.12 & -0.37\\
 $M_{K_s}$ & -1.79 & -1.65 &  -2.11 & -1.71 & -0.14 & -0.40\\
 \hline
  \multicolumn{1}{l|}{$N$}  &  \multicolumn{2}{c|}{45,877} & \multicolumn{2}{c|}{342,502} & \multicolumn{2}{c|}{} \\
\hline
\hline
\label{table:Table3}
\end{tabular}
\end{table}

\section{RC Contamination}
RC stars are metal rich horizontal branch stars, that span a narrow range on $\log g$ and a relatively wide range on $T_{eff}$. In this study, the most probable RC stars are selected using multiple Gaussian distributions on $T_{eff}$ and $\log g$ parameters. RC selection is done by selecting stars within the 2$\sigma$ region around the central coordinates of these distributions. In Fig. \ref{fig:Fig2} there is an evident peak of RC on the $\log g$ distribution and there is a large overlapping region with RGB stars below of this distribution. On the contrary, the RC population appears reasonably well separated in the distribution on the $T_{eff}$. Based on these distributions, the RC sample seems to be contaminated with a small percentage of sub-giants and RGB stars, but these sub-samples reside in 3$\sigma$ region of the central coordinates of $\log g \times T_{eff}$ distribution.


Contamination of AGRC sample is also tested using the empirical method of \citet{Holtzman18}, the method that considers the [C/N] ratio as a proxy for mass \citep{Martig16, Ness15} to identify RGB stars. In this method, a reference temperature is calculated with their Eq.~10. Then, $T_{eff}$ and $\log g$ are linked to [C/N] abundance ratio in order to assign the stellar population using Eq.~11 of \citet{Holtzman18}. In this study, this method is only applied to the APOGEE RC stars, because APOGEE survey provides both carbon and nitrogen abundances while GALAH survey provides only carbon. So, this separation method is only applied to the APOGEE giants. Based on the analysis of 10,760 RC giants, the contamination of RGB stars is found 0.35\%. Instead,  the contamination levels of GALAH RC stars are unknown.

In recent years \citet{Bovy14} provided a new catalogue of RC stars compiled from APOGEE DR11 by applying a new technique on spectrophotometric data. This catalogue is further updated in APOGEE DR15. This catalogue have 5\% contamination.

Our AGRC sample coincides with the 65\% of Bovy's RC catalogue of APOGEE DR15. The remaining 35\% fraction is missing because of (i) application of relative parallax error limitation to our sample ($\sigma_{\varpi} /\varpi \leq 0.1$, (ii) Differences between stellar atmospheric model parameters arising because the original RC catalogue of \citet{Bovy14}, that uses APOGEE DR11, is extended to the DR15 data release. The average difference in $T_{eff}$ is 55~K with $\sigma_{T_{eff}}$ = 45~K while, he average difference in logarithmic surface gravity is 0.1~dex with $\sigma_{\log~g}$ = 0.15~dex. These differences can explain why only 65\% of the stars matched.

\section{Summary and Conclusion} \label{sec:style}
In this study, absolute magnitudes and colours of RC stars selected from high resolution and high $S/N$ APOGEE DR14 \citep{Majewski17} and GALAH DR2 \citep{dS16} spectroscopic surveys, are examined in a range of UV, optical, NIR and MIR photometric bands of {\it GALEX} GR6/7 \citep{Bianchi17}, SDSS DR7 \citep{Abazijan09}, {\it Gaia} DR2 \citep{GC18}, 2MASS \citep{Cutri03} and All-{\it WISE} \citep{Cutri13} photometric surveys, respectively. In this study, RC stars are not used as standard candles. Instead, their standard candle value is thoroughly examined. The RC star distances are individually estimated with 1/$\varpi$ and BJ18 method by using {\it Gaia} DR2 parallaxes.

This study provides a new method to select RC stars using overlapping Gaussian distributions on the spectroscopic plane. Also, the separation of RC stars into the chemical populations is done by running a machine learning algorithm that produces a decision boundary line to provide a clear separation on chemical plane. RC stars are identified using the Gaussian distributions of each contributing parameter of the HR diagram. Initially sources with $\sigma_{\varpi}/\varpi \leq 0.1$ are selected to obtain more accurate astrometric sample. This procedure is followed by the application of the GMM method to separate RC stars into low-$\alpha$ (thin disc) and high-$\alpha$ (thick disc). By doing so, chemical space solves most of the Galactic population separation problem, which is the best choice so far and this approach is never used before in the literature. RC sample is further cross-matched with {\it GALEX} GR6/7, SDSS DR7, {\it Gaia} DR2, 2MASS and All-{\it WISE} catalogues and all of the sub-samples are evaluated in their own unique conditions, which in turn is given the non-unique number of sources in each photometric survey. 

Results are evaluated for each photometric band and colour (i) by comparing absolute magnitude values in low- and high-$\alpha$ populations, (ii) by comparing different distance estimation methods, and (iii) by comparing observational results for 2MASS $J$, $H$ and $K_s$ bands with a mock catalogue generated for the current known parameters of our Galaxy.

The overall comparison of the absolute magnitudes and colours with the systematic analyses that is used in this study indicates that there exist specific trends between low- and high-$\alpha$ populations, regardless of the distance estimation method. A general trend of brightening in the UV region of the electromagnetic spectrum is observed, while a trend of fainting in optical and infrared bands is found going from low-$\alpha$ to high-$\alpha$ populations. For $1/\varpi$, absolute magnitudes of RC stars vary between +0.12 and -0.13 mag from low- to high-$\alpha$ populations in UV to MIR band. Similarly, for BJ18's method, these variations are +0.13 and -0.17 mag in the same electromagnetic region. Brightening in UV bands are an expected behaviour for the older Galactic component so that this component corresponds to the high-$\alpha$ population or so called chemical thick disc. Thick disc stars are more metal poor than thin disc ones, causing generally higher fluxes in the UV bands of the former component. However, it was interesting to observe decreasing behaviour in absolute magnitudes from optical to infrared. These findings imply that in optic, NIR and MIR regions $\alpha$-elements might be contributing to the opacity in the stellar atmosphere. In short, the high-$\alpha$ AGRC stars are bluer in UV and become redder as the wavelength range of the chosen photometric band move towards the MIR bands (see Table \ref{table:Table4}).

Based on the comparison of observational and model results given in Table \ref{table:Table3}, a similar trend of decreasing absolute magnitudes from low-$\alpha$ to high-$\alpha$ population is confirmed from the mock data. However, differences in absolute magnitudes for the mock catalogue are around three times larger than the observational results.

Even though {\it Gaia} trigonometric parallax measurements reach down to faint magnitudes ($G=21$ mag), their relative parallax errors are increasing with apparent magnitudes. Estimated distances using photometric parallax method with absolute magnitude values of chemical population will allow more precise results than {\it Gaia} trigonometric parallax measurements of faint RC stars.

\citet{Martin19}, investigated the shape of the Galactic bulge using RC stars from the NIR Vista-VVV survey. In the study, they showed that the density distribution of RC stars towards the Bulge has two peaks. They claim that there is a 0.40 mag difference between these peaks, and the RC population cannot be represented with a single absolute magnitude. Similar behaviour that is presented in this study also confirms this absolute magnitude change in RC populations. These are the signatures of the chemical evolution of the Galaxy.

Observational data of RC stars have a wide range of use. RC stars are used as probes in astronomy by measuring distances, extinction, density, age, kinematics and chemical evolution probes of the Galaxy. This clear variation of RC stars properties  with chemical composition indicates that all the relations regarding the above investigations require re-consideration.

Our overall result is that the RC absolute magnitudes are depending on chemical populations independently from metallicity and distance estimation methods. This study gives explicit confirmation of the chemical population dependence of RC stars.

\section{acknowledgements}

We would like to thank the anonymous referee for his/her numerous contribution towards improving the Paper. This work has been supported by the Scientific and Technological Research Council of Turkey (T\"UB\. ITAK), Grant No: MFAG-118F350. This study is partially related to Olcay Plevne's PhD thesis.

We have also made use of data from: (1) the APOGEE survey, which is part of Sloan Digital Sky Survey IV. SDSS-IV is managed by the Astrophysical Research Consortium for the Participating Institutions of the SDSS Collaboration (http: //www.sdss.org). 

The GALAH survey is based on observations made at the Australian Astronomical Observatory, under programmes A/2013B/13, A/2014A/25, A/2015A/19, A/2017A/18. We acknowledge the traditional owners of the land on which the AAT stands, the Gamilaraay people, and pay our respects to elders past and present.

This research has also made use of data from the Sloan Digital Sky Survey (SDSS-DR7). 

This work has made use of data from the European Space Agency (ESA) mission {\it Gaia} (\url{https://www.cosmos.esa.int/gaia}), processed by the {\it Gaia}Data Processing and Analysis Consortium (DPAC, \url{https://www.cosmos.esa.int/web/gaia/dpac/consortium}). Funding for the DPAC has been provided by national institutions, in particular the institutions participating in the {\it Gaia} Multilateral Agreement. 

This publication makes use of data products from the Two Micron All Sky Survey, which is a joint project of the University of Massachusetts and the Infrared Processing and Analysis Center/California Institute of Technology, funded by the National Aeronautics and Space Administration and the National Science Foundation.

This publication makes use of data products from the Wide-field Infrared Survey Explorer, which is a joint project of the University of California, Los Angeles, and the Jet Propulsion Laboratory/California Institute of Technology, funded by the National Aeronautics and Space Administration.

The following software and programming languages made this research possible: Python (versions 2.7 \& 3.6); Astropy (version 2.0; Astropy Collaboration et al. 2013; The Astropy Collaboration et al. 2018), pandas (version 0.20.2; Mckinney 2011); This research has made use of the VizieR catalogue access tool, CDS, Strasbourg, France. The original description of the VizieR service was published in A\&AS 143, 23. For the analysis, the following software packages have been used: NumPy (Oliphant 2015), matplotlib (Hunter 2007), Jupyter Notebook (Kluyver et al. 2016).

\appendix

\setcounter{table}{0}
\renewcommand{\thetable}{A\arabic{table}}
\setcounter{figure}{0}
\renewcommand{\thefigure}{A\arabic{figure}}

\section{ABSOLUTE MAGNITUDES AND COLOURS FROM UV TO MIR}

\begin{figure}[h]
    \centering
    \includegraphics[width=\textwidth]{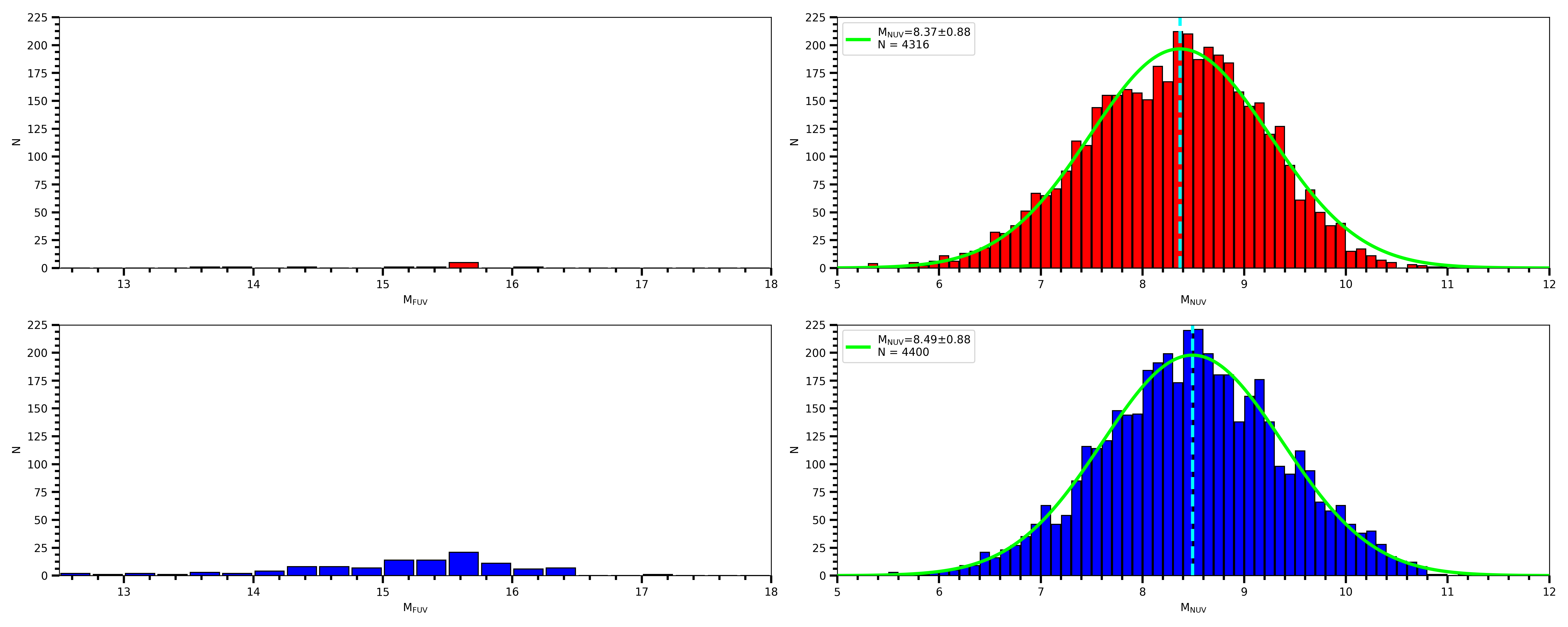}
\caption{Absolute magnitude distribution of RC stars in {\it GALEX} $NUV$ band that are calculated with the 1/$\varpi$ method for high-$\alpha$ (upper panel) and low-$\alpha$ (lower panel) populations. Green solid line is the Gaussian fit for the distribution and turquoise dashed line is the median value of the distribution. } 
\label{fig:Fig8}
\end {figure}  

\begin{figure}[h]
    \centering
    \includegraphics[width=\textwidth]{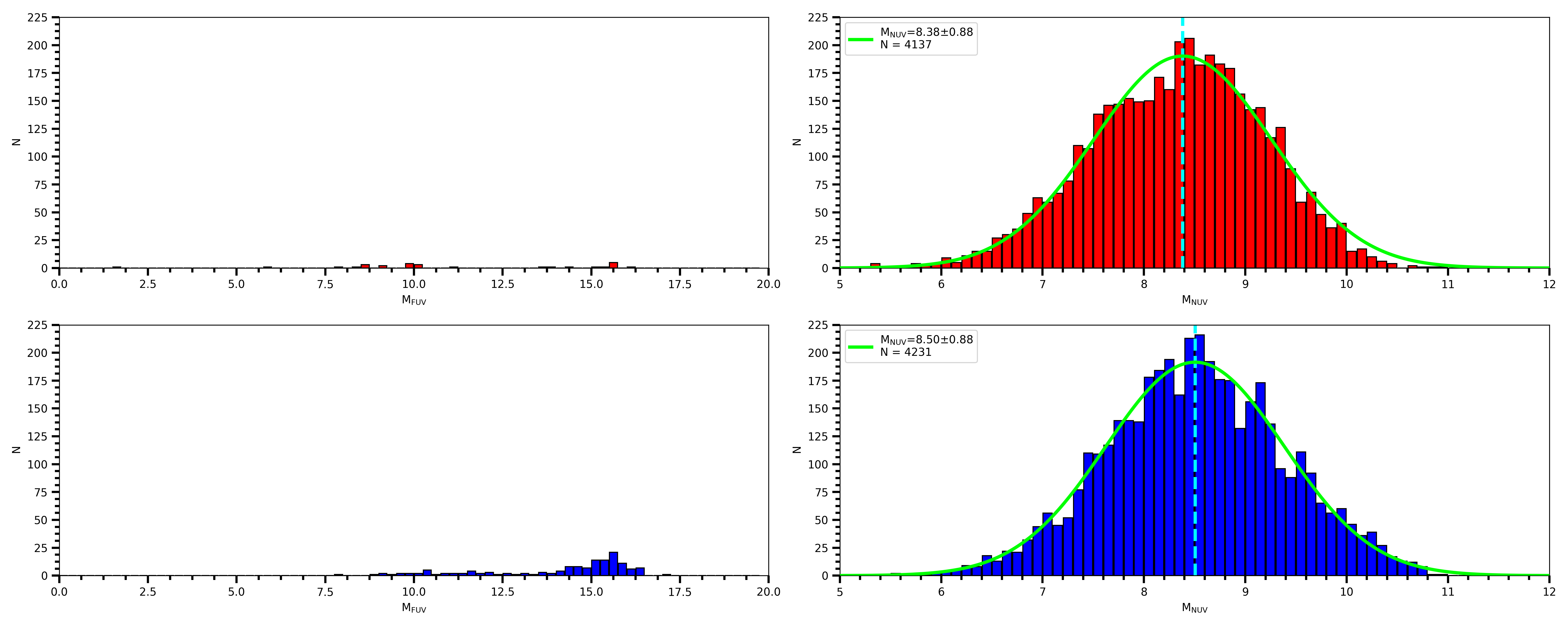}
\caption{Absolute magnitude distribution of RC stars in {\it GALEX} $NUV$ band that are calculated with the BJ18 method for high-$\alpha$ (upper panel) and low-$\alpha$ (lower panel) populations. Green solid line is the Gaussian fit for the distribution and turquoise dashed line is the median value of the distribution.} 
\label{fig:Fig9}
\end {figure}  

\begin{figure}[h]
    \centering
    \includegraphics[width=\textwidth]{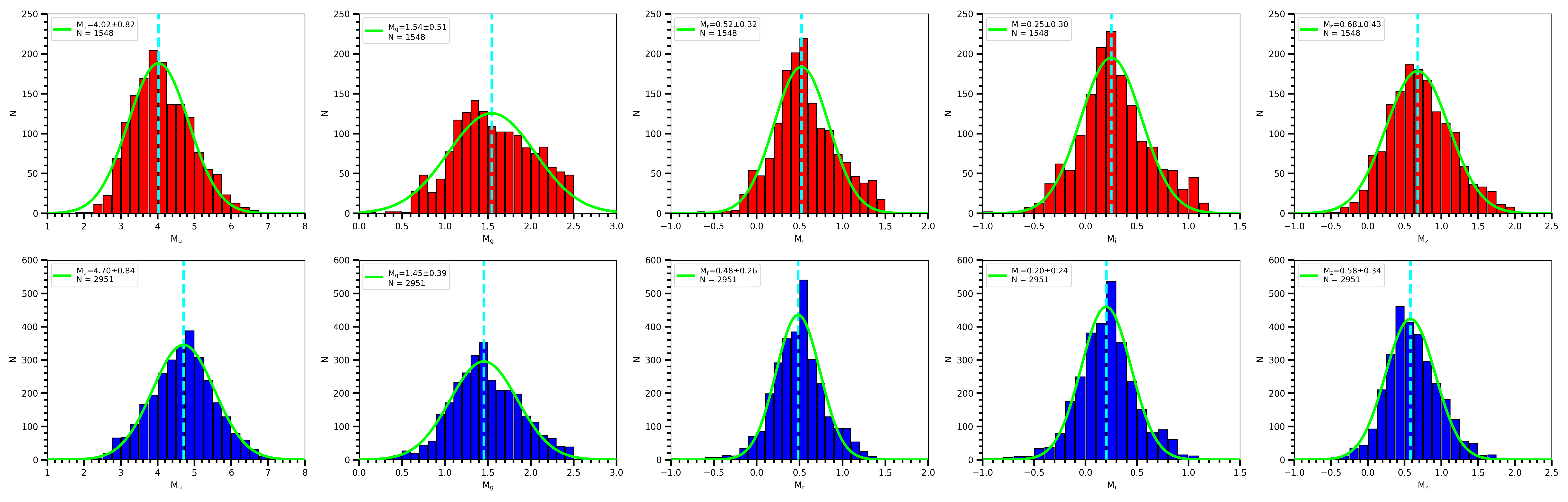}
\caption{Absolute magnitude distribution of RC stars in SDSS $u$, $g$, $r$, $i$ and $z$ bands that are calculated with the 1/$\varpi$ method for high-$\alpha$ (upper panel) and low-$\alpha$ (lower panel) populations. Green solid line is the Gaussian fit for the distribution and turquoise dashed line is the median value of the distribution.} 
\label{fig:Fig10}
\end {figure}  
 
\begin{figure}[h]
    \centering
    \includegraphics[width=\textwidth]{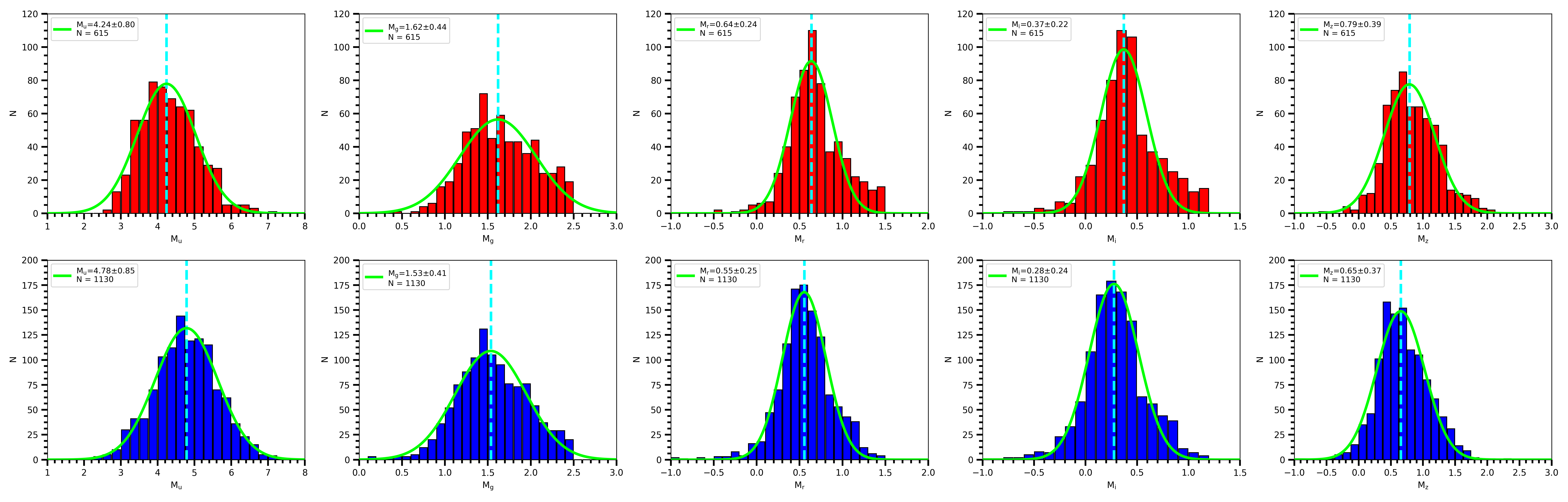}
\caption{Absolute magnitude distribution of RC stars in SDSS $u$, $g$, $r$, $i$ and $z$ bands that are calculated with the BJ18 method for high-$\alpha$ (upper panel) and low-$\alpha$ (lower panel) populations. Green solid line is the Gaussian fit for the distribution and turquoise dashed line is the median value of the distribution. } 
\label{fig:Fig11}
\end {figure}  

\begin{figure}[h]
    \centering
    \includegraphics[width=\textwidth]{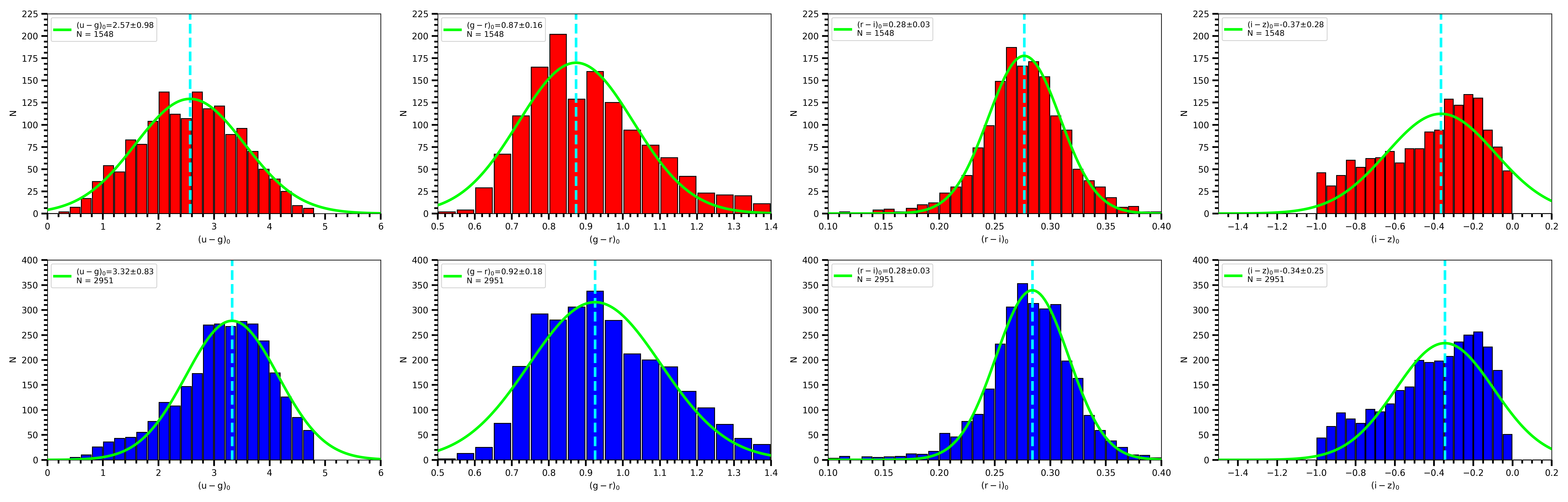}
\caption{Colour distribution of RC stars in SDSS $u-g$, $g-r$, $r-i$, and $i-z$ colours that are calculated with the 1/$\varpi$ method for high-$\alpha$ (upper panel) and low-$\alpha$ (lower panel) populations. Green solid line is the Gaussian fit for the distribution and turquoise dashed line is the median value of the distribution. } 
\label{fig:Fig12}
\end {figure}  

\begin{figure}[h]
    \centering
    \includegraphics[width=\textwidth]{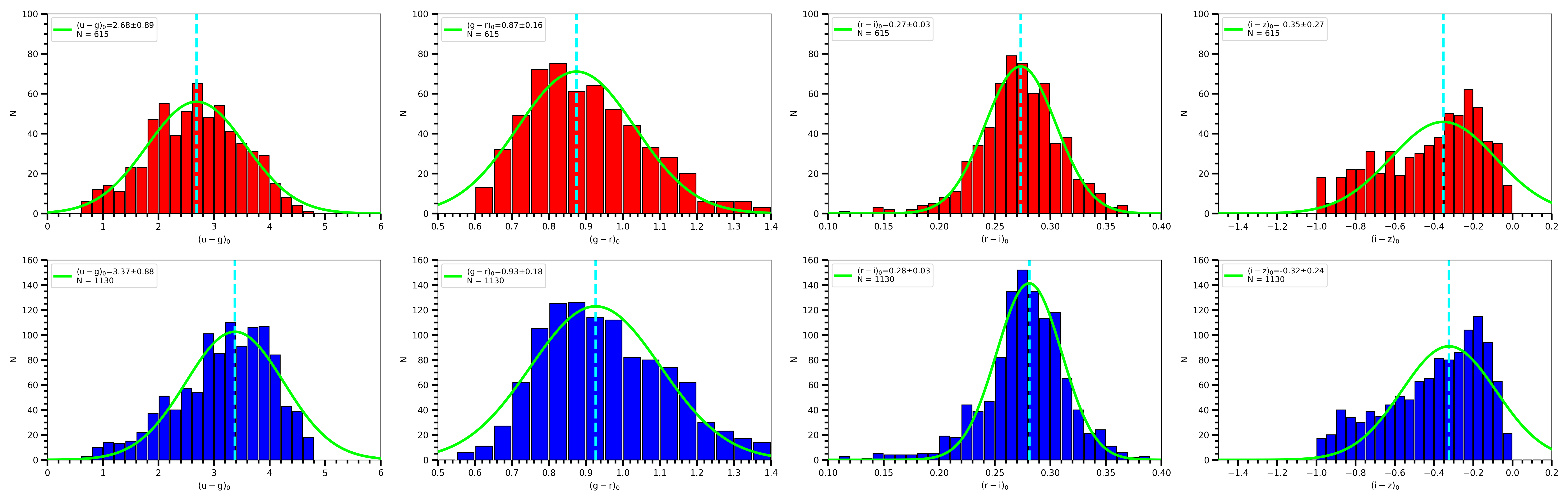}
\caption{Colour distribution of RC stars in SDSS $u-g$, $g-r$, $r-i$, and $i-z$ colours that are calculated with the BJ18 method forhigh-$\alpha$ (upper panel) and low-$\alpha$ (lower panel) populations. Green solid line is the Gaussian fit for the distribution and turquoise dashed line is the median value of the distribution. } 
\label{fig:Fig13}
\end {figure}  

\begin{figure}[h]
    \centering
    \includegraphics[width=\textwidth]{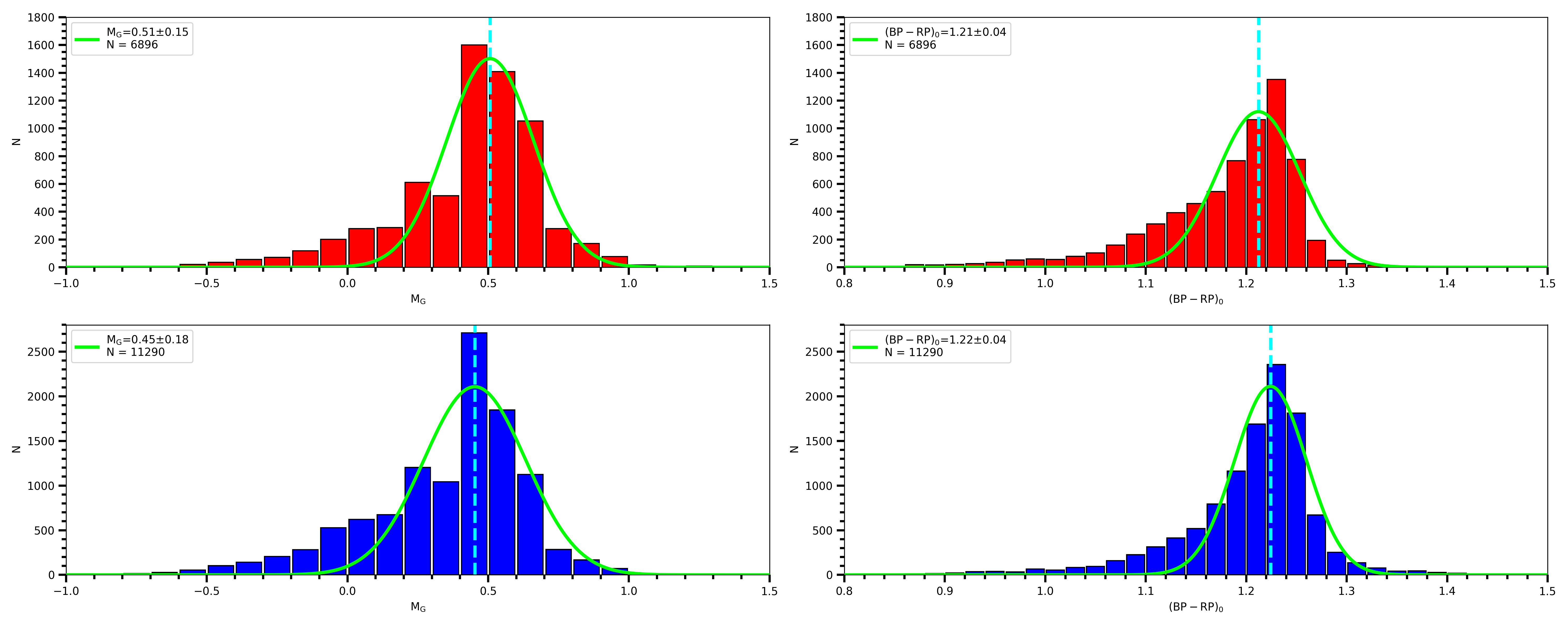}
\caption{Absolute magnitude distribution of RC stars in {\it Gaia} $G$ band and $G_{BP}-G_{RP}$ colour that are calculated with the 1/$\varpi$ method for high-$\alpha$ (upper panel) and low-$\alpha$ (lower panel) populations. Green solid line is the Gaussian fit for the distribution and turquoise dashed line is the median value of the distribution.} 
\label{fig:Fig14}
\end {figure}  

\begin{figure}[h]
    \centering
    \includegraphics[width=\textwidth]{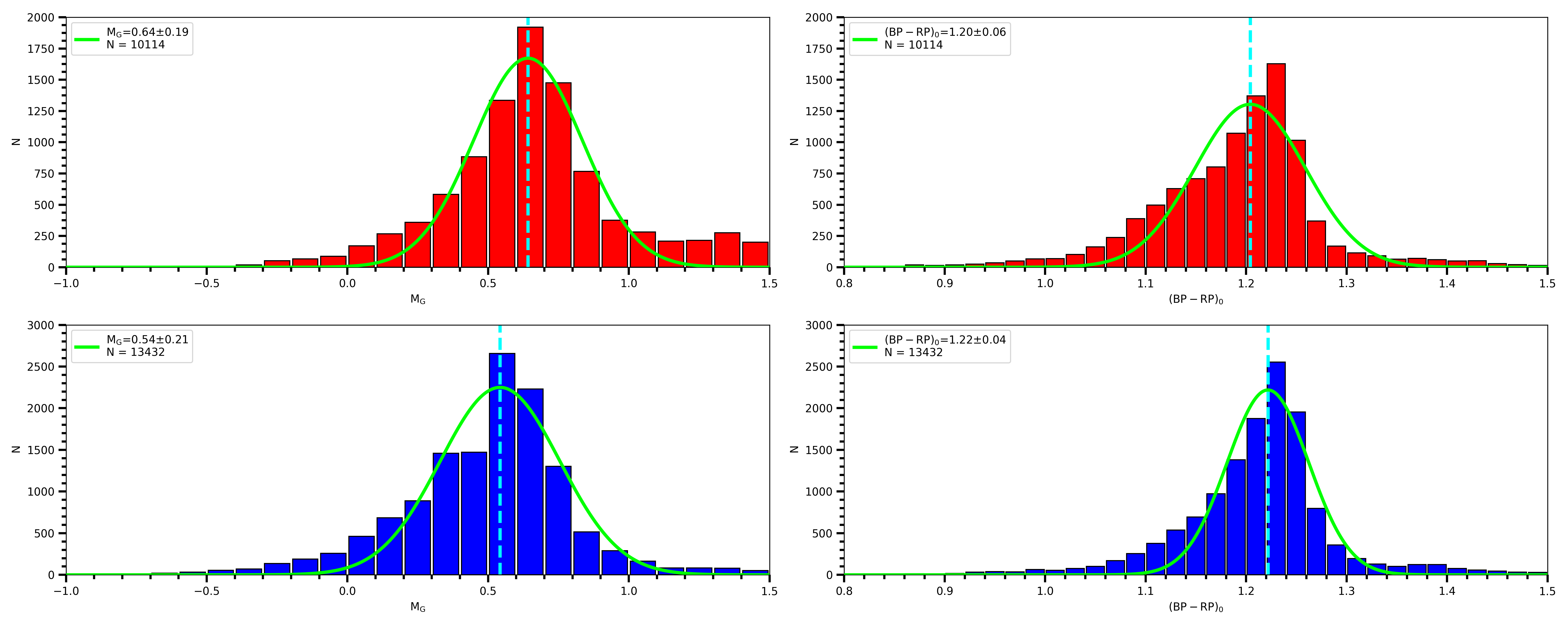}
\caption{Absolute magnitude distribution of RC stars in {\it Gaia} $G$ band and $G_{BP}-G_{RP}$ colour that are calculated with the BJ18 method for high-$\alpha$ (upper panel) and low-$\alpha$ (lower panel) populations. Green solid line is the Gaussian fit for the distribution and turquoise dashed line is the median value of the distribution.} 
\label{fig:Fig15}
\end {figure}  

\begin{figure}[h]
    \centering
    \includegraphics[width=\textwidth]{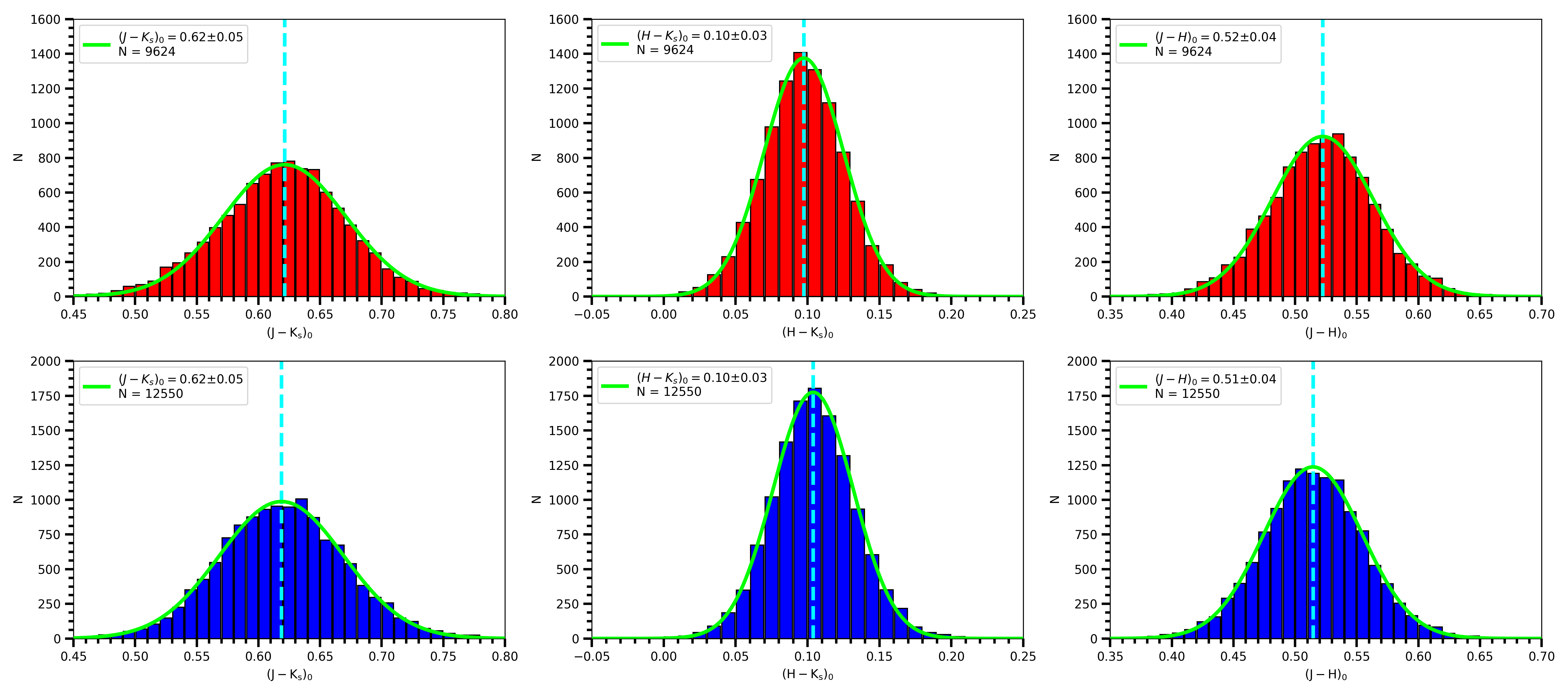}
\caption{Colour distribution of RC stars in 2MASS $J-K_s$, $H-K_s$, and $J-H$ colours that are calculated with the 1/$\varpi$ method for high-$\alpha$ (upper panel) and low-$\alpha$ (lower panel) populations. Green solid line is the Gaussian fit for the distribution and turquoise dashed line is the median value of the distribution. } 
\label{fig:Fig18}
\end {figure}  

\begin{figure}[h]
    \centering
    \includegraphics[width=\textwidth]{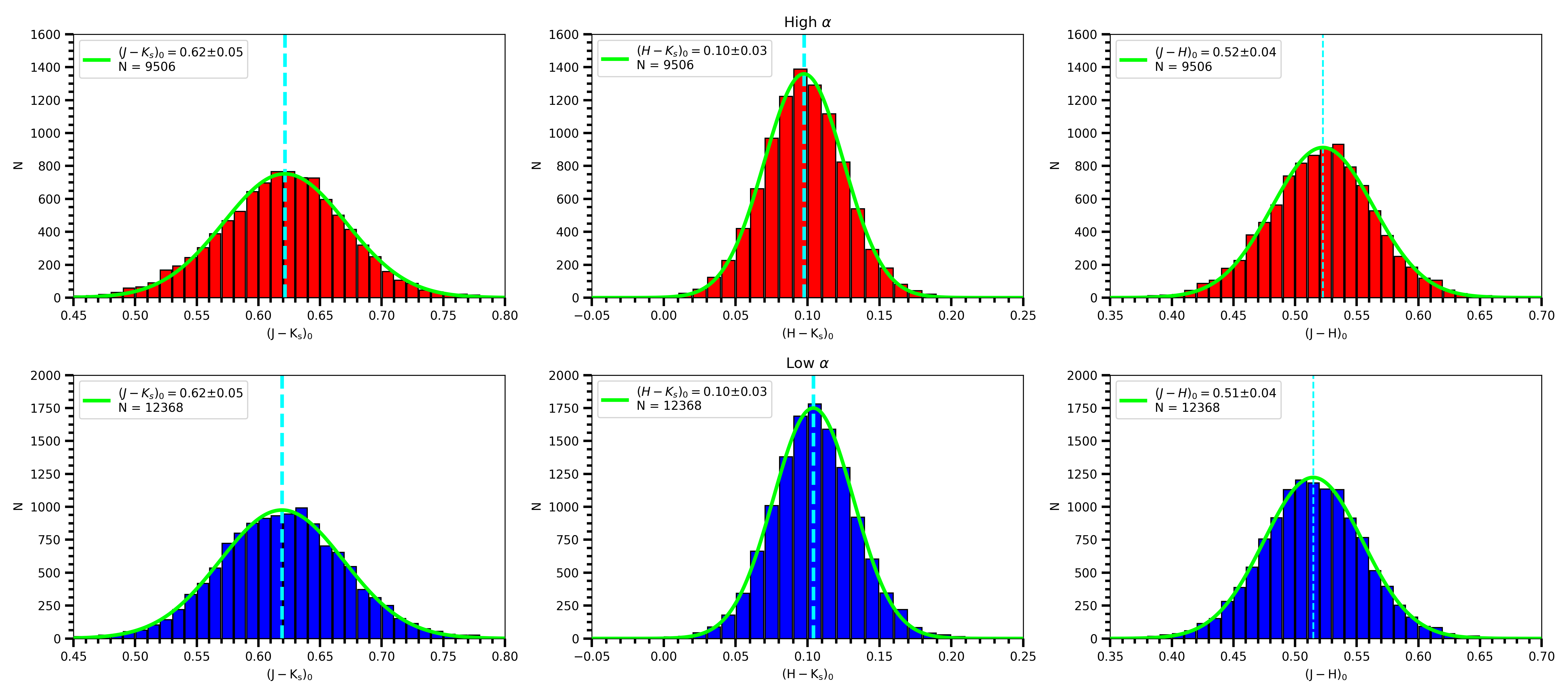}
\caption{Colour distribution of RC stars  in 2MASS $J-K_s$, $H-K_s$, and $J-H$ colours that are calculated with the BJ18 method for high-$\alpha$ (upper panel) and low-$\alpha$ (lower panel) populations. Green solid line is the Gaussian fit for the distribution and turquoise dashed line is the median value of the distribution. } 
\label{fig:Fig19}
\end {figure}  

\begin{figure}[h]
    \centering
    \includegraphics[width=\textwidth]{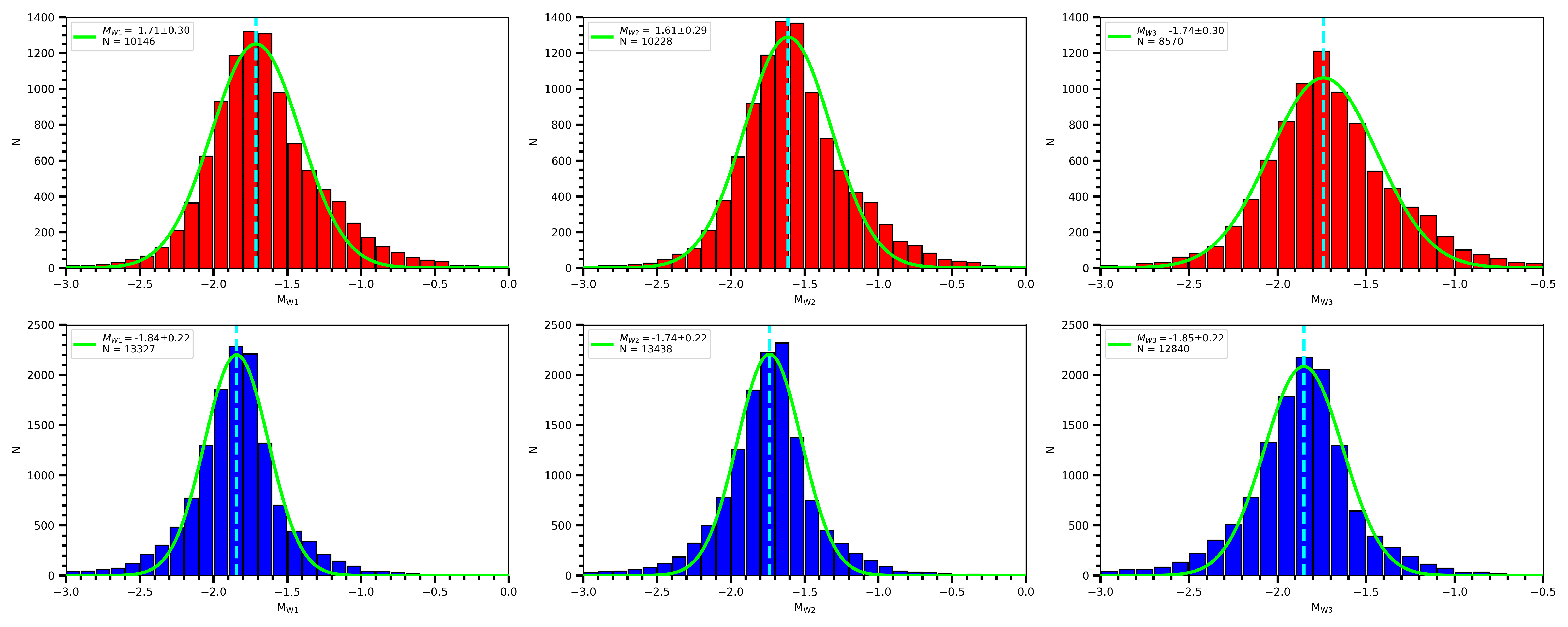}
\caption{Absolute magnitude distribution of RC stars in $W1$, $W2$, $W3$ and $W4$ bands that are calculated with the 1/$\varpi$ method for high-$\alpha$ (upper panel) and low-$\alpha$ (lower panel) populations.} 
\label{fig:Fig20}
\end {figure}  
 
\begin{figure}[h]
    \centering
    \includegraphics[width=\textwidth]{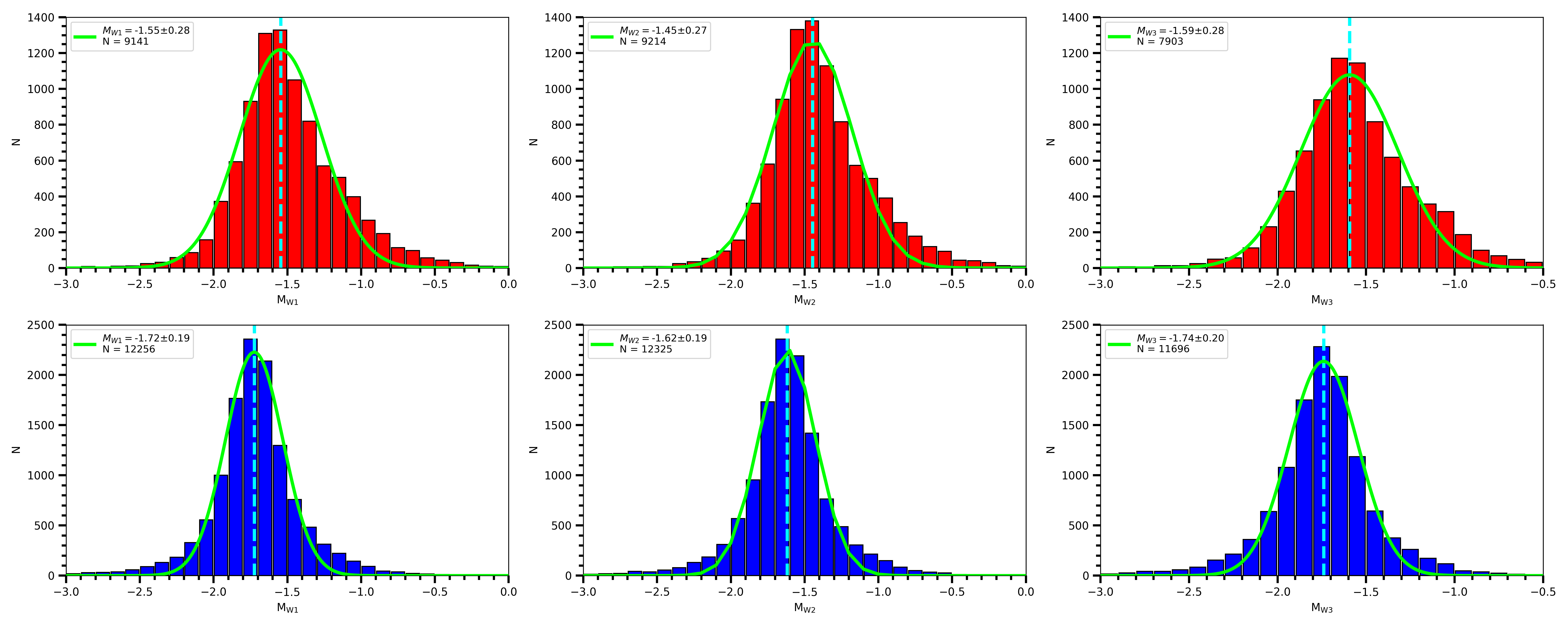}
\caption{Absolute magnitude distribution of RC stars in $W1$, $W2$, $W3$ and $W4$ bands that are calculated with the BJ18 method for high-$\alpha$ (upper panel) and low-$\alpha$ (lower panel) populations. Green solid line is the Gaussian fit for the distribution and turquoise dashed line is the median value of the distribution. } 
\label{fig:Fig21}
\end {figure}  

\begin{figure}[h]
    \centering
    \includegraphics[width=\textwidth]{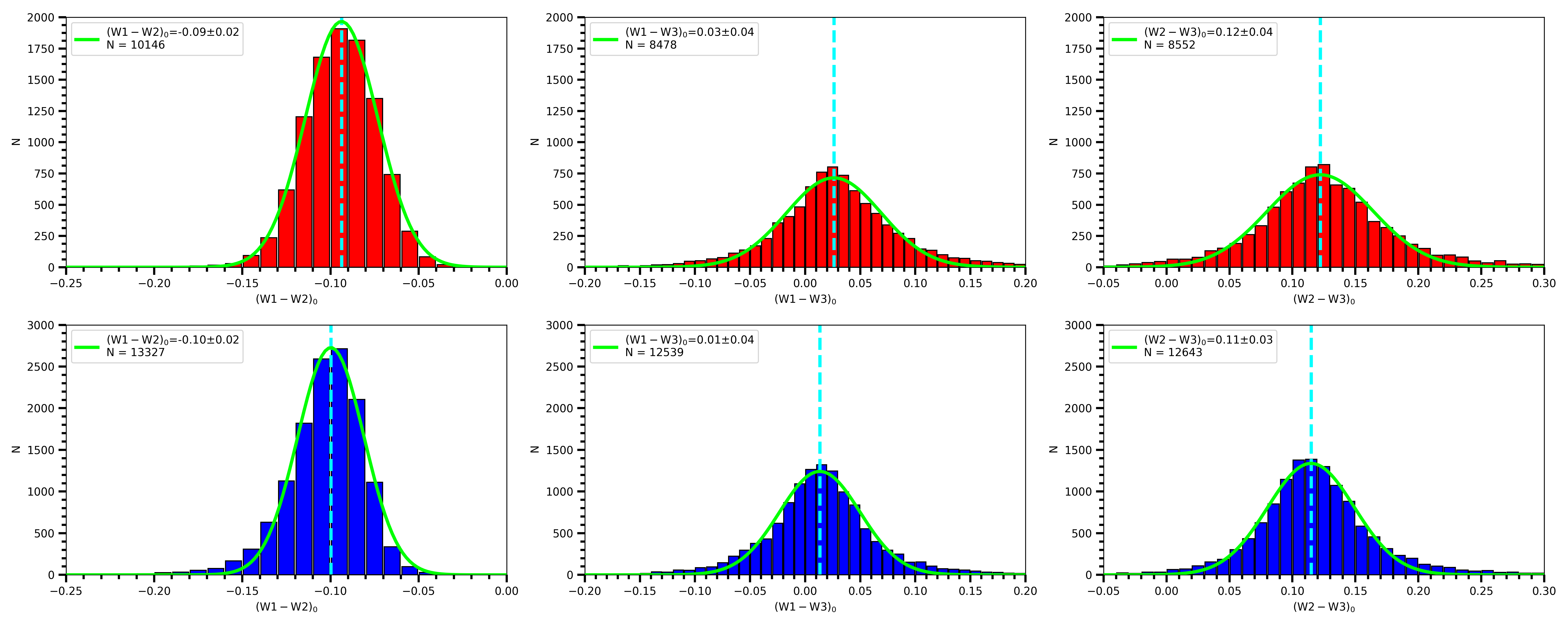}
\caption{Colour distribution of RC stars in $W1-W2$, $W1-W3$, and $W2-W3$ colours that are calculated with the 1/$\varpi$ method for high-$\alpha$ (upper panel) and low-$\alpha$ (lower panel) populations. Green solid line is the Gaussian fit for the distribution and turquoise dashed line is the median value of the distribution. } 
\label{fig:Fig22}
\end {figure}  

\begin{figure}[h]
    \centering
    \includegraphics[width=\textwidth]{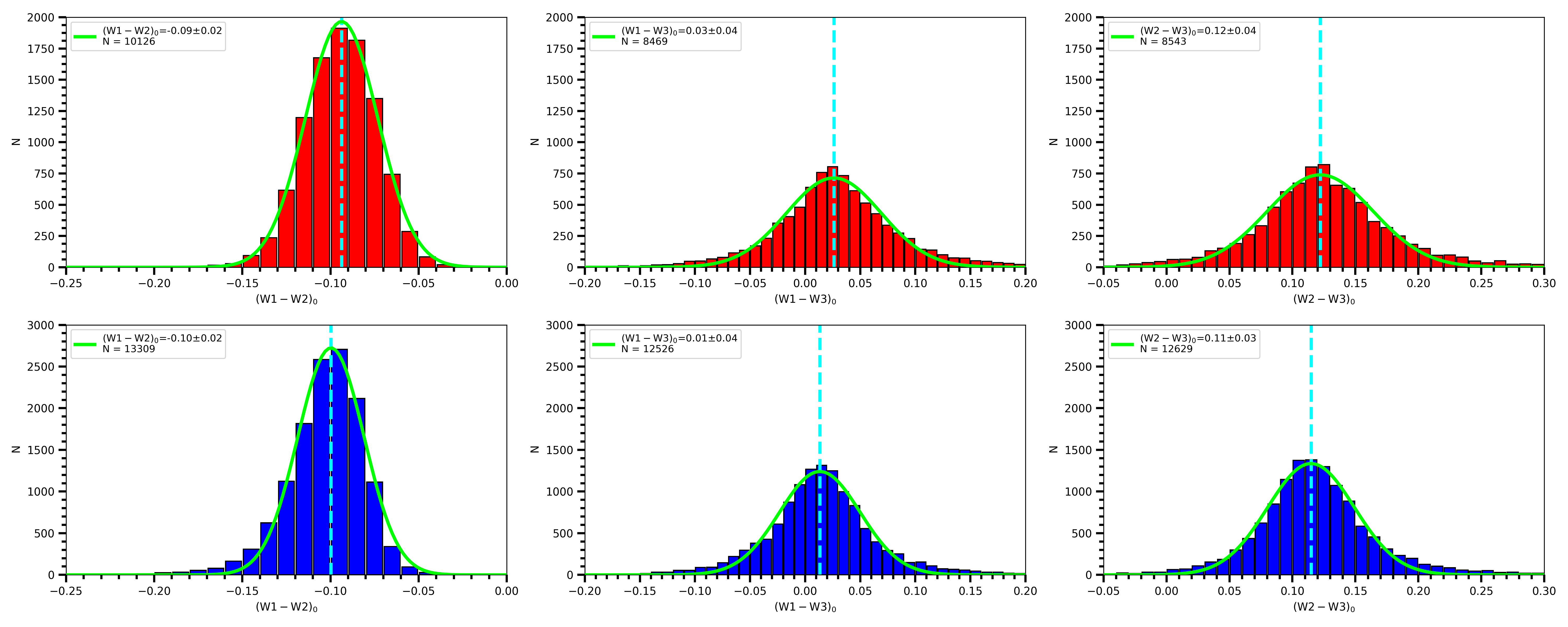}
\caption{Colour distribution of RC stars in $W1-W2$, $W1-W3$, and $W2-W3$ colours that are calculated with the BJ18 method for high-$\alpha$ (upper panel) and low-$\alpha$ (lower panel) populations. Green solid line is the Gaussian fit for the distribution and turquoise dashed line is the median value of the distribution. } 
\label{fig:Fig23}
\end {figure}  

\end{document}